\documentstyle[seceq,preprint]{ptptex}

\newcommand{\bsigma}{\mbox{{\boldmath $\sigma$}}}
\newcommand{\lsim}{~\ooalign{\hfil{\mbox{\Large${}_{\sim}$}}\hfil\crcr{\mbox{\raisebox{0.4ex}{$<$}}}}~}
\newcommand{\gsim}{~\ooalign{\hfil{\mbox{\Large${}_{\sim}$}}\hfil\crcr{\mbox{\raisebox{0.4ex}{$>$}}}}~}
\preprintnumber{NUP-A-97-4}

\title{Fine-Tuning Problem in a Left-Right Symmetric Model and Sogami's Generalized Covariant Derivative Method}

\author{Eizou  UMEZAWA\footnote{E-mail address: umezawa@phys.cst.nihon-u.ac.jp}}

\inst{Department of Physics, College of Science and Technology\\
Nihon University, Tokyo 101}

\recdate{
\today
}

\abst{We study Sogami's generalized covariant derivative method for a $SU(2)_L \times SU(2)_R \times U(1)_{B-L} \times SU(3)_{c}$ model that contains bi-doublet and triplet Higgs bosons. In particular, a detailed study is made on the minimization conditions of the Higgs potential. It is known that a minimization condition and certain phenomenology for a extra gauge boson mass derive a restriction on potential parameters, which requires fine-tuning. We show that the restriction can be reduced to a condition of Yukawa coupling constants giving a heavy mass of the right-handed tau neutrino in our model. We also discuss the consistency among parameter restrictions of our model by taking phenomenology of the Higgs boson masses into account.}

\begin{document}

\maketitle
\section{Introduction}
Recently, Sogami proposed a method constructing a lagrangian of the standard model with the aid of generalized covariant derivatives that contain Higgs fields in addition to gauge fields.\cite{1} \ Subsequently, Morita studied a way to introduce parameters into such a covariant derivative method in a systematic manner; there, he also studied the applicability of Sogami's method to non-gauge theories.\cite{2} \ Sogami's method has a close relation with the non-commutative geometric (NCG) method for gauge theories proposed by Connes and Lott,\cite{3} \ to which many reformulations have been attempted by many authors\cite{4} because of its unfamiliar geometrical structure. From the viewpoint of practical applications, Sogami's method is more useful than the NCG method; we can apply this method to some models without depending on geometrical concepts.\cite{2,5}

In this paper, we study a $SU(2)_L \times SU(2)_R \times U(1)_{B-L} \times SU(3)_{c}$ left-right (L-R) symmetric model based on Sogami and Morita's method. Our model contains three kinds of Higgs fields, a bi-doublet of $SU(2)_L \times SU(2)_R$, a triplet of $SU(2)_{L}$ and a triplet of $SU(2)_{R}$. 

The Higgs potential of our model must be invariant under the Lorentz, gauge and L-R transformations. The most general form of such a potential has a complicated structure. Fine-tuning of the potential parameters is necessary for the potential to be at a minimum when all the Higgs fields are evaluated at their respective vacuum expectation values (VEV) that are consistent with a certain phenomenology.\cite{6} \ On the other hand, Sogami's method gives rise to some constraints among coupling constants of a gauge-Higgs lagrangian, e.\,g., some relations between the Higgs potential parameters and Yukawa coupling constants. Such constraints among the coupling constants may affect the fine-tuning problem of the L-R symmetric model. From this point of view, we study the minimization conditions of the Higgs potential. We also discuss the consistency among parameter restrictions of our model, including restrictions arising from phenomenology of the Higgs boson masses.

In the next section, we review the conventional L-R symmetric model and the fine-tuning problem of this model. In \S 3, we construct a L-R symmetric model according to Sogami and Morita's method, and discuss the fine-tuning problem in our model. This section contains the main result of this paper. Section 4 consists of three subsections: In \S 4.1, we discuss the consistency among parameter restrictions, apart from the Higgs boson mass phenomenology. In \S 4.2, we give mass eigenvalues of the Higgs bosons. We consider in \S 4.3 the parameter tuning under all the restrictions. Section 5 is devoted to conclusion.
\section{Brief review of a $SU(2)_L \times SU(2)_R \times U(1)_{B-L} \times SU(3)_{c}$ model and a fine-tuning problem}
In the $SU(2)_L \times SU(2)_R \times U(1)_{B-L} \times SU(3)_{c}$ model, lepton and quark fields belong to $SU(2)_{L}$ or $SU(2)_{R}$ doublets according to their chirality. We write each field as
\begin{equation}
l_{L,R}^i =\pmatrix{\nu_{L,R}^i \cr e_{L,R}^i},~~~q_{L,R}^i =\pmatrix{u_{L,R}^i \cr d_{L,R}^i},\label{matter}
\end{equation}
where each component of the quark fields belongs to $SU(3)_{c}$ triplets, and indices $i=1\sim 3$ denote the generation. These fields also have $U(1)_{B-L}$ quantum numbers, where $B$ and $L$ denote the baryon and lepton numbers, respectively. 

Gauge bosons for the gauge symmetries $SU(2)_L$, $SU(2)_R$, $U(1)_{B-L}$ and $SU(3)_{c}$ are written as $W^{L}_{\mu a}$, $W^{R}_{\mu a}$, $B_{\mu}$ and $G_{\mu a}$, respectively. We use the notation
\begin{equation}
{\bf W}_{\mu}^{L}=W_{\mu a}^{L}\frac{{\bsigma}_{a}}{2},~{\bf W}_{\mu}^{R}=W_{\mu a}^{R}\frac{{\bsigma}_{a}}{2},~B_{\mu},~G_{\mu}=G_{\mu a}\frac{\lambda_{a}}{2},
\end{equation}
where ${\bsigma}_{a}~ (a=1,2,3)$ and $\lambda_{a}~ (a=1\sim 8)$ are, respectively, Pauli and Gell-Mann matrices normalized so that $tr({\bsigma}_{a}{\bsigma}_{b})=2\delta_{ab}$ and $tr(\lambda_{a}\lambda_{b})=2\delta_{ab}$. We use boldface for $2\times 2$ matrices acting on $SU(2)$ doublets.

To yield small masses for conventional neutrinos and heavy masses for right-handed neutrinos through the seesaw mechanism, it is preferable to introduce the Higgs fields
\begin{equation}
{\bf \Phi}= \pmatrix{\phi_{1}^{0}&\phi_{1}^{+}\cr \phi_{2}^{-}&\phi_{2}^{0}\cr },~~~
{\bf \Delta}_{L,R}= \pmatrix{\delta_{L,R}^{+}/\sqrt{2}&\delta_{L,R}^{++}\cr \delta_{L,R}^{0}&-\delta_{L,R}^{+}/\sqrt{2}\cr },
\end{equation}
where ${\bf \Phi}$ is a bi-doublet of $SU(2)_L \times SU(2)_R$ and ${\bf \Delta}_{L,R}$ is a triplet of $SU(2)_{L,R}$.\cite{6,7} \ The $U(1)_{B-L}$ quantum numbers of ${\bf \Phi}$ and ${\bf \Delta}_{L,R}$ are $0$ and $2$, respectively. These fields transform according to
\begin{eqnarray}
{\bf \Delta}_{L,R}&\to &(U_{B-L}^{l\ast})^{2}{\bf U_{L,R}}{\bf \Delta}_{L,R}{\bf U_{L,R}}^{\dag},\label{gauge trans. pro.1}\\
{\bf \Phi}&\to &{\bf U_{L}}{\bf \Phi}{\bf U_{R}}^{\dag},\label{gauge trans. pro.2}
\end{eqnarray}
where ${\bf U}_{L,R}$ ($U_{B-L}^{l}$) is an element of the $SU(2)$ ($U(1)$) transformations acting on the doublets (lepton fields).

The L-R transformation of this model is defined by
\begin{equation}
l_{L}^{i}\leftrightarrow l_{R}^{i}~,~~q_{L}^{i}\leftrightarrow q_{R}^{i}~,~~{\bf \Delta}_{L}\leftrightarrow {\bf \Delta}_{R}~,~~{\bf \Phi}\leftrightarrow {\bf \Phi}^{\dag}.
\end{equation}
The lagrangian of the model must be invariant under the gauge transformation and the L-R transformation. The general lagrangian density for the fermion fields is
\begin{eqnarray}
{\cal L}_{F~\mbox{{\footnotesize general}}}&=&{\bar l}_{L}i({\ooalign{\hfil/\hfil\crcr $\partial$}}+ig_{1}\frac{1}{2}{\ooalign{\hfil/\hfil\crcr $B$}}-ig_{2}{\ooalign{\hfil/\hfil\crcr ${\bf W}$}}^{L})l_{L}~+(L~\leftrightarrow ~R)\nonumber \\
&&+{\bar q}_{L}i({\ooalign{\hfil/\hfil\crcr $\partial$}}-ig_{1}\frac{1}{6}{\ooalign{\hfil/\hfil\crcr $B$}}-ig_{2}{\ooalign{\hfil/\hfil\crcr ${\bf W}$}}^{L}-ig_{3}{\ooalign{\hfil/\hfil\crcr $G$}})q_{L}~+(L~\leftrightarrow ~R)\label{fermionic lagrangian}\nonumber \\
&&+{\cal L}_{Y},\label{general Lf}
\end{eqnarray}
where
\begin{eqnarray}
{\cal L}_{Y}&=&-\{({\bar l}_{L} h^{l}_{g}{\bf \Phi}l_{R} +{\bar l}_{L} {\tilde h}^{l}_{g} {\tilde {\bf \Phi}}l_{R}+{\bar q}_{L} h^{q}_{g}{\bf \Phi}q_{R} +{\bar q}_{L} {\tilde h}^{q}_{g} {\tilde {\bf \Phi}}q_{R}) ~+(\mbox{h.c.})\}\nonumber \\
&&-\{(il_{L}^{T}C{\bsigma}_{2}f_{g}{\bf \Delta}_{L}l_{L}+il_{R}^{T}C{\bsigma}_{2}f_{g}{\bf \Delta}_{R}l_{R}) +~(\mbox{h.c.})\}\label{general Lf-Y}
\end{eqnarray}
and ${\tilde {\bf \Phi}}=(i{\bsigma}_{2}){\bf \Phi}^{\ast}(i{\bsigma}_{2})^{\dag}$. Here, $g_{k}~(k=1 \sim 3)$ are the gauge coupling constants of $U(1)_{B-L}$, $SU(2)_{L,R}$ and $SU(3)_{c}$, respectively, and $f_{g}$, $h^{l,q}_{g}$ and ${\tilde h}^{l,q}_{g}$ are fermion-Higgs Yukawa coupling matrices having generation indices; $h^{l,q}_{g}$ and ${\tilde h}^{l,q}_{g}$ must be hermitian for the L-R symmetry of the fermionic part of the lagrangian. For simplicity, we have omitted matrix indices representing the generation. The general lagrangian density for boson fields is
\begin{eqnarray}
{\cal L}_{B~\mbox{{\footnotesize general}}}&=&-\frac{1}{4}(F_{\mu \nu}^{B}F^{B\mu \nu}+F_{\mu \nu a}^{L}F^{L\mu \nu a}+F_{\mu \nu a}^{R}F^{R\mu \nu a}+G_{\mu \nu}^{a}G^{a\mu \nu})\nonumber \\
&&+tr|D_{\mu}{\bf \Delta}_{L}|^{2}+tr|D_{\mu}{\bf \Delta}_{R}|^{2}+tr|D_{\mu}{\bf \Phi}|^{2}\nonumber \\
&&-V_{\mbox{{\footnotesize general}}},\label{general Lb}
\end{eqnarray}
where
\begin{eqnarray}
D_{\mu}{\bf \Delta}_{L,R}&=&\partial_{\mu}{\bf \Delta}_{L,R}-ig_{1}B_{\mu}{\bf \Delta}_{L,R}-ig_{2}[{\bf W}_{\mu}^{L,R},{\bf \Delta}_{L,R}],\\
D_{\mu}{\bf \Phi}&=&\partial_{\mu}{\bf \Phi}-ig_{2}({\bf W}_{\mu}^{L}{\bf \Phi}-{\bf \Phi}{\bf W}_{\mu}^{R})
\end{eqnarray}
and $F_{\mu \nu}^{B}$, $F_{\mu \nu a}^{L,R}$ and $G_{\mu \nu a}$ are field strengths of $B_{\mu}$, $W_{\mu a}^{L,R}$ and $G_{\mu a}$, respectively. Here, $V_{\mbox{{\footnotesize general}}}$ is the most general Higgs potential of this model consisting of linearly independent terms: according to Ref.~\citen{6}, we write this as
\begin{eqnarray}
V_{\mbox{{\footnotesize general}}}=&-&\mu_{1}^{2}tr({\bf \Phi}^{\dag}{\bf \Phi})-\mu_{2}^{2}\{tr({\tilde {\bf \Phi}}{\bf \Phi}^{\dag})+tr({\tilde {\bf \Phi}}^{\dag}{\bf \Phi})\}\nonumber \\
&-&\mu_{3}^{2}\{tr({\bf \Delta}_{L}{\bf \Delta}_{L}^{\dag})+tr({\bf \Delta}_{R}{\bf \Delta}_{R}^{\dag})\}\nonumber \\
&+&\lambda_{1}\{tr({\bf \Phi}^{\dag}{\bf \Phi})\}^{2}+\lambda_{2}[\{tr({\tilde {\bf \Phi}}{\bf \Phi}^{\dag})\}^{2}+\{tr({\tilde {\bf \Phi}}^{\dag}{\bf \Phi})\}^{2}]\nonumber \\
&+&\lambda_{3}tr({\tilde {\bf \Phi}}{\bf \Phi}^{\dag})tr({\tilde {\bf \Phi}}^{\dag}{\bf \Phi})\nonumber \\
&+&\lambda_{4}tr({\bf \Phi}^{\dag}{\bf \Phi})\{tr({\tilde {\bf \Phi}}{\bf \Phi}^{\dag})+tr({\tilde {\bf \Phi}}^{\dag}{\bf \Phi})\}\nonumber \\
&+&\rho_{1}[\{tr({\bf \Delta}_{L}{\bf \Delta}_{L}^{\dag})\}^{2}+\{tr({\bf \Delta}_{R}{\bf \Delta}_{R}^{\dag})\}^{2}]\nonumber \\
&+&\rho_{2}\{tr({\bf \Delta}_{L}{\bf \Delta}_{L})tr({\bf \Delta}_{L}^{\dag}{\bf \Delta}_{L}^{\dag})+tr({\bf \Delta}_{R}{\bf \Delta}_{R})tr({\bf \Delta}_{R}^{\dag}{\bf \Delta}_{R}^{\dag})\}\nonumber \\
&+&\rho_{3}tr({\bf \Delta}_{L}{\bf \Delta}_{L}^{\dag})tr({\bf \Delta}_{R}{\bf \Delta}_{R}^{\dag})\nonumber \\
&+&\rho_{4}\{tr({\bf \Delta}_{L}{\bf \Delta}_{L})tr({\bf \Delta}_{R}^{\dag}{\bf \Delta}_{R}^{\dag})+tr({\bf \Delta}_{L}^{\dag}{\bf \Delta}_{L}^{\dag})tr({\bf \Delta}_{R}{\bf \Delta}_{R})\}\nonumber \\
&+&\alpha_{1}tr({\bf \Phi}^{\dag}{\bf \Phi})\{tr({\bf \Delta}_{L}{\bf \Delta}_{L}^{\dag})+tr({\bf \Delta}_{R}{\bf \Delta}_{R}^{\dag})\}\nonumber \\
&+&\alpha_{2}\{tr({\bf \Phi}{\tilde {\bf \Phi}}^{\dag})tr({\bf \Delta}_{R}{\bf \Delta}_{R}^{\dag})+tr({\bf \Phi}^{\dag}{\tilde {\bf \Phi}})tr({\bf \Delta}_{L}{\bf \Delta}_{L}^{\dag})\}\nonumber \\
&+&\alpha_{2}^{\ast}\{tr({\bf \Phi}^{\dag}{\tilde {\bf \Phi}})tr({\bf \Delta}_{R}{\bf \Delta}_{R}^{\dag})+tr({\bf \Phi}{\tilde {\bf \Phi}}^{\dag})tr({\bf \Delta}_{L}{\bf \Delta}_{L}^{\dag})\}\nonumber \\
&+&\alpha_{3}\{tr({\bf \Phi}{\bf \Phi}^{\dag}{\bf \Delta}_{L}{\bf \Delta}_{L}^{\dag})+tr({\bf \Phi}^{\dag}{\bf \Phi}{\bf \Delta}_{R}{\bf \Delta}_{R}^{\dag})\}\nonumber \\
&+&\beta_{1}\{tr({\bf \Phi}{\bf \Delta}_{R}{\bf \Phi}^{\dag}{\bf \Delta}_{L}^{\dag})+tr({\bf \Phi}^{\dag}{\bf \Delta}_{L}{\bf \Phi}{\bf \Delta}_{R}^{\dag})\}\nonumber \\
&+&\beta_{2}\{tr({\tilde {\bf \Phi}}{\bf \Delta}_{R}{\bf \Phi}^{\dag}{\bf \Delta}_{L}^{\dag})+tr({\tilde {\bf \Phi}}^{\dag}{\bf \Delta}_{L}{\bf \Phi}{\bf \Delta}_{R}^{\dag})\}\nonumber \\
&+&\beta_{3}\{tr({\bf \Phi}{\bf \Delta}_{R}{\tilde {\bf \Phi}}^{\dag}{\bf \Delta}_{L}^{\dag})+tr({\bf \Phi}^{\dag}{\bf \Delta}_{L}{\tilde {\bf \Phi}}{\bf \Delta}_{R}^{\dag})\}.\label{general V}
\end{eqnarray}
Here the coefficients other than $\alpha_{2}$ must be real to make the lagrangian hermitian and L-R invariant.

Non-zero VEV of neutral Higgs fields break the gauge symmetry down to $U(1)_{EM}$. We write the VEV of the Higgs fields as
\begin{equation} 
<{\bf \Delta}_{L,R}>_{0}=\frac{1}{\sqrt{2}}\pmatrix{0&0\cr v_{L,R}&0\cr },~~~
<{\bf \Phi}>_{0}=\frac{1}{\sqrt{2}}\pmatrix{\epsilon_{1}&0\cr 0&\epsilon_{2}\cr }.\label{VEVs}
\end{equation}
To yield a phenomenologically desired mass for each particle,
\begin{equation}
|v_{L}|\ll |\epsilon_{1,2}|\ll |v_{R}|\label{seesaw}
\end{equation}
is required.\cite{6,7} \ Mass expressions of fermions and charged gauge bosons that are determined by Eqs.~(\ref{general Lf}), (\ref{general Lb}), (\ref{VEVs}) and (\ref{seesaw}) are
\begin{eqnarray}
m_{e^{i}}&\simeq &\frac{1}{2\sqrt{2}}\left|h^{l}_{i}\epsilon_{2}+{\tilde h}_{i}^{l}\epsilon_{1}\right|,\label{e}\\
m_{u^{i}}&\simeq &\frac{1}{2\sqrt{2}}\left|h^{q}_{i}\epsilon_{1}+{\tilde h}_{i}^{q}\epsilon_{2}\right|,\label{u}\\
m_{d^{i}}&\simeq &\frac{1}{2\sqrt{2}}\left|h^{q}_{i}\epsilon_{2}+{\tilde h}_{i}^{q}\epsilon_{1}\right|,\label{d}\\
m_{\nu_{R}^{i}}&\simeq &\frac{1}{\sqrt{2}}\left|f_{i}v_{R}\right|,\label{lepton mass-1}\\
m_{\nu_{L}^{i}}&\simeq &\frac{1}{\sqrt{2}}\left|f_{i}v_{L}-\frac{(h_{Di})^{2}\epsilon_{+}^{2}}{2f_{i}v_{R}}\right|,\label{lepton mass-2}\\
m_{W_{L}}^{2}&\simeq &\frac{1}{4}g_{2}^{2}\epsilon_{+}^{2},\label{WL mass}\\
m_{W_{R}}^{2}&\simeq &\frac{1}{2}g_{2}^{2}v_{R}^{2},\label{WR mass}
\end{eqnarray}
where we have used the definitions
\begin{equation}
f=2f_{g},~h^{l,q}=2h^{l,q}_{g},~({\tilde h}^{l,q})^{T}=2{\tilde h}^{l,q}_{g}\label{redef}
\end{equation}
and $f_{i}=(f)_{ii}$, $h^{l,q}_{i}=(h^{l,q})_{ii}$, ${\tilde h}^{l,q}_{i}=({\tilde h}^{l,q})_{ii}$, $h_{Di}=(h^{l}_{i}\epsilon_{1}+{\tilde h}^{l}_{i}\epsilon_{2})/\sqrt{2}\epsilon_{+}$ and $\epsilon_{\pm}^{2}=\epsilon_{1}^{2}\pm \epsilon_{2}^{2}$, while ignoring the generation mixing. From this point, we refer to $f$, $h^{l,q}$ and ${\tilde h}^{l,q}$ as Yukawa coupling matrices.

Next, we review a fine-tuning problem of the L-R symmetric model, which is derived from a minimization condition of the Higgs potential and a phenomenological restriction on the VEV of the Higgs fields. In general, if we choose $\alpha_{2}$, which is the only complex coupling in the general Higgs potential, to be real, we can make all the VEV of the Higgs fields real. In this case we obtain four non-trivial extremal conditions that determine a minimal point of the potential:\cite{6}
\begin{eqnarray}
\mu_{1}^{2}&=&\frac{1}{2\epsilon_{-}^{2}}\{2v_{L}v_{R}(\beta_{2}\epsilon_{1}^{2}-\beta_{3}\epsilon_{2}^{2})+(v_{L}^{2}+v_{R}^{2})(\alpha_{1}\epsilon_{-}^{2}-\alpha_{3}\epsilon_{2}^{2})\}\nonumber \\
&&+\epsilon_{+}^{2}\lambda_{1}+2\epsilon_{1}\epsilon_{2}\lambda_{4},\label{1-1}\\
\mu_{2}^{2}&=&\frac{1}{4\epsilon_{-}^{2}}[v_{L}v_{R}\{\beta_{1}\epsilon_{-}^{2}-2\epsilon_{1}\epsilon_{2}(\beta_{2}-\beta_{3})\}+(v_{L}^{2}+v_{R}^{2})(2\alpha_{2}\epsilon_{-}^{2}+\alpha_{3}\epsilon_{1}\epsilon_{2})]\nonumber \\
&&+\epsilon_{1}\epsilon_{2}(2\lambda_{2}+\lambda_{3})+\frac{\lambda_{4}\epsilon_{+}^{2}}{2},\label{1-2}\\
\mu_{3}^{2}&=&\frac{1}{2}\{\alpha_{1}\epsilon_{+}^{2}+4\alpha_{2}\epsilon_{1}\epsilon_{2}+\alpha_{3}\epsilon_{2}^{2}+2\rho_{1}(v_{L}^{2}+v_{R}^{2})\},\label{1-3}\\
\beta_{2}&=&\frac{1}{\epsilon_{1}^{2}}\{-\beta_{1}\epsilon_{1}\epsilon_{2}-\beta_{3}\epsilon_{2}^{2}+(2\rho_{1}-\rho_{3})v_{L}v_{R}\}.\label{1-4}
\end{eqnarray}
It is known that a serious problem comes from Eq.~(\ref{1-4}), which can be rewritten as
\begin{equation}
\frac{\beta_{1}\epsilon_{1}\epsilon_{2}+\beta_{2}\epsilon_{1}^{2}+\beta_{3}\epsilon_{2}^{2}}{(2\rho _{1}-\rho _{3})\epsilon_{+}^{2}}=\frac{v_{L}v_{R}}{\epsilon_{+}^{2}},\label{minimization-1}
\end{equation}
where the $\beta_{i}$ and the $\rho_{i}$ are quartic Higgs self-coupling constants. If we set $m_{W_{R}}$ to $\simeq 1.4\,\mbox{TeV}$ as a mass in an experimentally accessible energy range in the near future,\cite{6} \ the value on the right-hand side of this equation is restricted to
\begin{equation}
\left|\frac{v_{L}v_{R}}{\epsilon_{+}^{2}}-\frac{h_{D1}^{2}}{2f_{1}^{2}}\right| \simeq \frac{m_{\nu_{L}^{1}}m_{W_{R}}^{2}}{2m_{\nu_{R}^{1}}m_{W_{L}}^{2}}~~<~~10^{-8}.\label{phenomenological restriction}
\end{equation}
To derive this, we have used Eqs.~(\ref{e}) $\sim$ (\ref{lepton mass-2}) with
\begin{eqnarray}
m_{\nu_{L}^{1}}&\lsim &10\,\mbox{eV},\\
m_{W_{L}}&\simeq &80\,\mbox{GeV},\\
m_{\nu_{R}^{1}}&>& 63\,\mbox{GeV}\left(\frac{1.6\,\mbox{TeV}}{m_{W_{R}}}\right)^{4},\label{phnuR1}
\end{eqnarray}
where inequality (\ref{phnuR1}) arises from an experimental limit on the neutrinoless double-$\beta$ decay.\cite{6,double} \ Substituting Eq.~(\ref{minimization-1}) into (\ref{phenomenological restriction}), we obtain a restriction on the Higgs self-coupling constants and the Yukawa coupling constants,
\begin{equation}
\left|\frac{\beta_{1}\epsilon_{1}\epsilon_{2}+\beta_{2}\epsilon_{1}^{2}+\beta_{3}\epsilon_{2}^{2}}{(2\rho _{1}-\rho _{3})\epsilon_{+}^{2}}-\frac{h_{D1}^{2}}{2f_{1}^{2}}\right| ~~<~~10^{-8}.\label{restriction of quartic couplings}
\end{equation}
Barring a highly tuned cancellation, this restriction requires both $\rho_{1},\rho_{3} \gg \beta_{1}, \beta_{2}, \beta_{3}$ and $2f_{1}^{2}\gg h_{D1}^{2}$. Such hierarchical choices of the coupling constants are unnatural.
\section{$SU(2)_L \times SU(2)_R \times U(1)_{B-L} \times SU(3)_{c}$ model based on Sogami's method}
To apply Sogami's method to the L-R symmetric model, we use the following representations of fermion fields:
\begin{equation}
\Psi^{li}=\pmatrix{l_L ^i , & l_L ^{ic}, & l_R ^i , & l_R ^{ic} \cr}^{T},~~~l_{L,R}^{ic}=i\gamma^2 l_{L,R}^{i\ast},
\end{equation}
\begin{equation}
\Psi^{qi}=\pmatrix{q_L ^i , & q_L ^{ic}, & q_R ^i , & q_R ^{ic} \cr}^{T},~~~q_{L,R}^{ic}=i\gamma^2 q_{L,R}^{i\ast},
\end{equation}
where $\{\gamma_{\mu},\gamma_{\nu}\}=2g_{\mu \nu}$ and $g_{\mu \nu}=\mbox{diag}(1,-1,-1,-1)$. We define generalized covariant derivatives in our model as operators acting on these fields:
\begin{eqnarray}
D_{\mu}^{lij}&=&\delta^{ij}\partial _{\mu}-i\delta^{ij}\sum _k ^{1,2}g_k A_{\mu}^{l(k)}+\frac{i}{4}\gamma _{\mu}A^{l(0)ij},\label{general Dl}\\
D_{\mu}^{qij}&=&\delta^{ij}\partial _{\mu}-i\delta^{ij}\sum _k ^{1,2,3}g_k A_{\mu}^{q(k)}+\frac{i}{4}\gamma _{\mu}A^{q(0)ij},\label{general Dq}
\end{eqnarray}
where
\begin{eqnarray}
A_{\mu}^{l(1)}&=&-\frac{1}{2}B_{\mu}\mbox{diag}({\bf 1},-{\bf 1},{\bf 1},-{\bf 1}),\\
A_{\mu}^{q(1)}&=&\frac{1}{6}B_{\mu}\mbox{diag}({\bf 1},-{\bf 1},{\bf 1},-{\bf 1}),\\
A_{\mu}^{l(2)}&=&A_{\mu}^{q(2)}=\mbox{diag}({\bf W}_{\mu}^{L},-{\bf W}_{\mu}^{L \ast},{\bf W}_{\mu}^{R},-{\bf W}_{\mu}^{R \ast}),\\
A_{\mu}^{q(3)}&=&\mbox{diag}(G_{\mu}{\bf 1},-G_{\mu}^{\ast}{\bf 1},G_{\mu}{\bf 1},-G_{\mu}^{\ast}{\bf 1}),\\
A^{l(0)ij}&=&{\cal E}\pmatrix{0&f^{\dag}{\bf \Delta}_L ^{\dag}&h^{l}{\bf \Phi}&0\cr f{\bf \Delta}_L &0&0&\tilde{h}^{l}{\bf \Phi}\cr h^{l}{\bf \Phi}^{\dag}&0&0&f^{\dag}{\bf \Delta}_R ^{\dag}\cr 0&\tilde{h}^{l}{\bf \Phi}^{\dag}&f{\bf \Delta}_R &0\cr}^{ij}\!\!{\cal E}^{\dag}+{\cal C}^{lij},\label{l-Higgs}\\
A_{\mu}^{q(0)ij}&=&{\cal E}\pmatrix{0&0&h^{q}{\bf \Phi}&0\cr 0&0&0&\tilde{h}^{q}{\bf \Phi}\cr h^{q}{\bf \Phi}^{\dag}&0&0&0\cr 0&\tilde{h}^{q}{\bf \Phi}^{\dag}&0&0\cr}^{ij}\!\!{\cal E}^{\dag}+{\cal C}^{qij}\label{q-Higgs}
\end{eqnarray}
and ${\cal E}=\mbox{diag}({\bf 1},i{\bsigma}_{2},{\bf 1},i{\bsigma}_{2})$. The quantities ${\cal C}^{l,q ij}$ are constant matrices. To make our bosonic lagrangian hermitian, we restrict their forms to 
\begin{eqnarray}
{\cal C}^{l ij}&=&\mbox{diag}(C_{P}^{l ij}{\bf 1},C_{A}^{l ij}{\bf 1},C_{P}^{l ij}{\bf 1},C_{A}^{l ij}{\bf 1}),\label{Cl}\\
{\cal C}^{q ij}&=&\mbox{diag}(C_{LP}^{q ij}{\bf 1},C_{LA}^{q ij}{\bf 1},C_{RP}^{q ij}{\bf 1},C_{RA}^{q ij}{\bf 1}),\label{C2}
\end{eqnarray}
where each constituent is real. The gauge transformations of $A_{\mu}^{l,q(k)} (k=1\sim 3)$ and $A^{l,q (0)ij}$ are defined by
\begin{eqnarray}
A_{\mu}^{l,q(k)}&\to&U^{l,q}A_{\mu}^{l,q(k)}U^{l,q\dag}+\frac{i}{g_k}U^{l,q(k)}\partial_{\mu}U^{l,q(k)\dag},\\
A^{l,q(0)ij}&\to&U^{l,q}A^{l,q(0)ij}U^{l,q\dag},
\end{eqnarray}
where
\begin{eqnarray}
U^{l}&=&U^{l(1)}U^{l(2)},\\
U^{q}&=&U^{q(1)}U^{q(2)}U^{q(3)}
\end{eqnarray}
and
\begin{eqnarray}
U^{l(1)}&=&\mbox{diag}(U_{B-L}^{l},U_{B-L}^{l \ast},U_{B-L}^{l},U_{B-L}^{l \ast}),\\
U^{q(1)}&=&\mbox{diag}(U_{B-L}^{q},U_{B-L}^{q \ast},U_{B-L}^{q},U_{B-L}^{q \ast}),\\
U^{l(2)}&=&U^{q(2)}=\mbox{diag}({\bf U}_{L},{\bf U}_{L}^{\ast},{\bf U}_{R},{\bf U}_{R}^{\ast}),\\
U^{q(3)}&=&\mbox{diag}(U_{c},U_{c}^{\ast},U_{c},U_{c}^{\ast})
\end{eqnarray}
and $U_{c}$ ($U_{B-L}^{q}$) is an element of the $SU(3)_{c}$ ($U(1)_{B-L}$) transformations acting on the quark fields.

The fermionic lagrangian density is defined by
\begin{equation}
{\cal L}_{F}=\sum_{X}^{l,q}\sum_{i,j}^{1\sim 3}\bar{\Psi} ^{X i}i\gamma^{\mu}D_{\mu}^{X ij}\Psi ^{X j}.\label{LF}
\end{equation}
Through an integration by parts, we obtain the general form of the fermionic lagrangian density of the L-R symmetric model that contains fermion-Higgs Yukawa interactions. (See Eqs.~(\ref{general Lf}), (\ref{general Lf-Y}) and (\ref{redef}).)

Generalized field strengths are defined by
\begin{equation}
\check{F}_{\mu \nu}^{l,q}=[D_{\mu}^{l,q},D_{\nu}^{l,q}],
\end{equation}
where we have omitted the matrix indices representing the generation. Their concrete forms are given in Appendix A.

Using $\check{F}_{\mu \nu}^{l,q}$, we define our bosonic lagrangian density as
\begin{eqnarray}
{\cal L}_{B}=-\frac{1}{4}\{&&\sum_{X}^{l,q} \sum_{t}^{S,V,A,T,P}\xi_{t}Tr(\check{F}_{\mu \nu}^{X}\Gamma_{t}E_{\bar{\delta}^{X}_{1}}\check{F}^{\mu \nu X}\Gamma^{t}E_{\delta^{X}_{1}})\nonumber \\
&&+\sum_{X}^{l,q} \sum_{t}^{S,V,A,T,P}\xi_{t}Tr(\check{F}_{\mu \nu}^{X}\sigma^{\mu \nu}\Gamma_{t}E_{\bar{\delta}^{X}_{2}}\check{F}_{\rho \sigma X}\sigma^{\rho \sigma }\Gamma^{t}E_{\delta^{X}_{2}})\nonumber \\
&&+\sum_{X}^{l,q}\frac{1}{6i}(TrE_{\eta^{X}_{1}}\sigma^{\mu \nu}\check{F}_{\mu \nu}^{X})\frac{1}{6i}(TrE_{\eta^{X}_{2}}\sigma^{\mu \nu}\check{F}_{\mu \nu}^{X})\nonumber \\
&&+\frac{1}{6i}(TrE_{\eta^{l}_{3}}\sigma^{\mu \nu}\check{F}_{\mu \nu}^{l})\frac{1}{6i}(TrE_{\eta^{q}_{3}}\sigma^{\mu \nu}\check{F}_{\mu \nu}^{q})\nonumber \\
&&+\sum_{X}^{l,q}\frac{1}{6i}(TrE_{\eta^{X}_{4}}\sigma^{\mu \nu}\check{F}_{\mu \nu}^{X})\},
\end{eqnarray}
where $Tr$ denotes the trace with respect to all matrix indices and $t=(S,V,A,T,P)$ corresponds to $\Gamma_{t}=(1_{c},\gamma_{\mu},\gamma^{5}\gamma_{\mu},\sigma_{\mu \nu},\gamma^{5})$, respectively.\footnote{$1_{c}$ denotes the unit matrix in the space of $4$-spinors and $\sigma_{\mu \nu}=i[\gamma_{\mu},\gamma_{\nu}]/2$. We also define $\Gamma^{t}=(1_{c},\gamma^{\mu},\gamma^{5}\gamma^{\mu},\sigma^{\mu \nu},\gamma^{5})$.} \ Further, the $\xi_{t}$ and
\begin{eqnarray}
E_{\delta^{X}_{a}}&=&\mbox{diag}~(\delta_{Pa}^{X}{\bf 1},\delta_{Aa}^{X}{\bf 1},\delta_{Pa}^{X}{\bf 1},\delta_{Aa}^{X}{\bf 1}), ~(a=1,2)\label{E1}\\
E_{\bar{\delta}^{X}_{a}}&=&\mbox{diag}~(\bar{\delta}_{Pa}^{X}{\bf 1},\bar{\delta}_{Aa}^{X}{\bf 1},\bar{\delta}_{Pa}^{X}{\bf 1},\bar{\delta}_{Aa}^{X}{\bf 1}), ~(a=1,2)\\
E_{\eta^{l}_{a}}&=&\mbox{diag}(\eta_{Pa}^{l}{\bf 1},\eta_{Aa}^{l}{\bf 1},\eta_{Pa}^{l}{\bf 1},\eta_{Aa}^{l}{\bf 1}), ~(a=1 \sim 4)\\
E_{\eta^{q}_{a}}&=&\mbox{diag}~(\eta_{LPa}^{q}{\bf 1},\eta_{LAa}^{q}{\bf 1},\eta_{RPa}^{l}{\bf 1},\eta_{RAa}^{q}{\bf 1})~(a=1 \sim 4)\label{E4}
\end{eqnarray}
are constants and constant matrices, respectively. To obtain ${\cal L}_{B}$ as a hermitian invariant under the gauge and L-R transformation, we assume that these matrices are real and hermitian, respectively. Note that, in general, the $\delta$, the $\bar{\delta}$ and the $\eta$ in Eqs.~(\ref{E1}) $\sim $ (\ref{E4}) may have matrix structures for the generation indices. We shall assume that the $\delta$ and the $\bar{\delta}$ are proportional to the unit matrix in \S 3. According to Ref.~\citen{2}, we also define the new parameters $\alpha$, $\beta$ and $\kappa$ related to the following summations: 
\begin{equation}
\sum_{t}\xi _{t} \Gamma _{t}\sigma_{\mu \nu}\Gamma^{t}=\alpha \sigma _{\mu \nu},~~
\sum_{t}\xi _{t} \Gamma _{t}\gamma_{\mu}\Gamma^{t}=\beta \gamma _{\mu},~~
\sum_{t}\xi _{t} \Gamma _{t}1_{c}\Gamma^{t}=\kappa 1_{c},
\end{equation}
where we do not require the positivity of these parameters. In the construction of ${\cal L}_{B}$, for generality, we have added several terms of $\sigma^{\mu \nu}\check{F}_{\mu \nu}^{X}$, which have an effect on the Higgs potential.

To obtain the correct coefficients of the kinetic terms of the gauge and Higgs fields, we require the following conditions:
\begin{eqnarray}
-\frac{1}{4}&=&g_{1}^{2}\{\kappa tr(\delta^{l}_{P1}\bar{\delta}^{l}_{P1}+\delta^{l}_{A1}\bar{\delta}^{l}_{A1}+\frac{1}{9}\delta^{q}_{P1}\bar{\delta}^{q}_{P1}+\frac{1}{9}\delta^{q}_{A1}\bar{\delta}^{q}_{A1})\nonumber \\
&&+2\alpha tr(\delta^{l}_{P2}\bar{\delta}^{l}_{P2}+\delta^{l}_{A2}\bar{\delta}^{l}_{A2}+\frac{1}{9}\delta^{q}_{P2}\bar{\delta}^{q}_{P2}+\frac{1}{9}\delta^{q}_{A2}\bar{\delta}^{q}_{A2})\},\label{1}\\
-\frac{1}{4}&=&\frac{g_{2}^{2}}{2}\{\kappa tr(\delta_{P1}^{l}\bar{\delta}^{l}_{P1}+\delta_{A1}^{l}\bar{\delta}^{l}_{A1}+\delta_{P1}^{q}\bar{\delta}^{q}_{P1}+\delta_{A1}^{q}\bar{\delta}^{q}_{A1})\nonumber \\
&&+2\alpha tr(\delta_{P2}^{l}\bar{\delta}^{l}_{P2}+\delta_{A2}^{l}\bar{\delta}^{l}_{A2}+\delta_{P2}^{q}\bar{\delta}^{q}_{P2}+\delta_{A2}^{q}\bar{\delta}^{q}_{A2})\},\label{2}\\
-\frac{1}{4}&=&2g_{3}^{2}\{\kappa tr(\delta_{P1}^{q}\bar{\delta}^{q}_{P1}+\delta_{A1}^{q}\bar{\delta}^{q}_{A1})+2\alpha tr(\delta_{P2}^{q}\bar{\delta}^{q}_{P2}+\delta_{A2}^{q}\bar{\delta}^{q}_{A2})\},\label{3}\\
\frac{1}{\beta}&=&\frac{3}{4}tr(h^{l}\bar{\delta}_{P1}^{l}h^{l}\delta_{P1}^{l}+{\tilde h}^{l}\bar{\delta}_{A1}^{l}{\tilde h}^{l}\delta_{A1}^{l}+h^{q}\bar{\delta}_{P1}^{q}h^{q}\delta_{P1}^{q}+{\tilde h}^{q}\bar{\delta}_{A1}^{q}{\tilde h}^{q}\delta_{A1}^{q})\nonumber \\
&&-\frac{9}{2}tr(h^{l}\bar{\delta}_{P2}^{l}h^{l}\delta_{P2}^{l}+{\tilde h}^{l}\bar{\delta}_{A2}^{l}{\tilde h}^{l}\delta_{A2}^{l}+h^{q}\bar{\delta}_{P2}^{q}h^{q}\delta_{P2}^{q}+{\tilde h}^{q}\bar{\delta}_{A2}^{q}{\tilde h}^{q}\delta_{A2}^{q}),\label{4}\\
\frac{1}{\beta}&=&\frac{3}{8}tr(f\bar{\delta}_{P1}^{l}f^{\dag}\delta_{A1}^{l}+f^{\dag}\bar{\delta}_{A1}^{l}f\delta_{P1}^{l})-\frac{9}{4}tr(f\bar{\delta}_{P2}^{l}f^{\dag}\delta_{A2}^{l}+f^{\dag}\bar{\delta}_{A2}^{l}f\delta_{P2}^{l}).~~\label{5}
\end{eqnarray}
Under these conditions, ${\cal L}_{B}$ have the same structure as the general form of the L-R symmetric model, except the Higgs potential. (See Eqs.~(\ref{general Lb}) and (\ref{general V}).) Our Higgs potential excludes several terms in comparison with the general form of the Higgs potential of Eq.~(\ref{general V}), i.\,e.,
\begin{equation}
\mu_{2}=\lambda_{2}=\lambda_{4}=\rho_{4}=\alpha{_2}=\beta_{2}=\beta_{3}=0.\label{zero}
\end{equation}
Non-zero coefficients are expressed in terms of the Yukawa coupling constants and several parameters introduced in the construction of ${\cal L}_{B}$. We give them in Appendix B. Our Higgs potential also contains a constant term that is excluded in Eq.~(\ref{general V}).

In the remaining part of this section, we consider the fine-tuning problem of the Higgs self-coupling constants in our model. In our model, the $\beta_{i}$ and the $\rho _{i}$ are expressed in terms of the Yukawa coupling constants and several parameters introduced in ${\cal L}_{B}$. If we assume the conditions
\begin{equation}
\delta_{Pa}^{l,q}=\delta_{Aa}^{l,q}\equiv \delta_{a}^{l,q}1_{g},~~
\bar{\delta}_{Pa}^{l,q}=\bar{\delta}_{Aa}^{l,q}\equiv \bar{\delta}_{a}^{l,q}1_{g},~(a=1,2)\label{tuning-1}
\end{equation}
where $1_{g}$ denotes the unit of the matrices for the generation indices and $\delta_{a}^{l,q}$ and $\bar{\delta}_{a}^{l,q}$ are constant numbers, we obtain
\begin{equation}
\frac{\beta_{1}\epsilon_{1}\epsilon_{2}+\beta_{2}\epsilon_{1}^{2}+\beta_{3}\epsilon_{2}^{2}}{(2\rho _{1}-\rho _{3})\epsilon_{+}^{2}}=\frac{\epsilon_{1}\epsilon_{2}}{\epsilon_{+}^{2}}\frac{tr(fh^{l}f^{\dag}\tilde{h}^{l})}{tr\{(ff^{\dag})^{2}\}}.
\end{equation}
This equation implies that in our model, Eqs.~(\ref{minimization-1}) and (\ref{phenomenological restriction}) yield a restriction on Yukawa coupling constants:
\begin{equation}
\left|\frac{\epsilon_{1}\epsilon_{2}}{\epsilon_{+}^{2}}\frac{tr(fh^{l}f^{\dag}\tilde{h}^{l})}{tr\{(ff^{\dag})^{2}\}}-\frac{h_{D1}^{2}}{2f_{1}^{2}}\right| < 10^{-8}.\label{main result}
\end{equation}
Therefore, under the simple conditions of Eq.~(\ref{tuning-1}), the fine-tuning of the Higgs self-coupling constants is reduced to the tuning of the Yukawa coupling constants. This is the main result of this paper. Note that the first and second term on the left-hand side of Eq.~(\ref{main result}) have the same structure, i.\,e., $(h_{i}^{l})^{2}/(f_{i})^{2}$, roughly speaking. Barring a highly tuned cancellation, this inequality requires $f_{i}\gg h^{l}_{i}, \tilde{h}^{l}_{i}$. As we shall discuss later, such a relation among the Yukawa coupling constants is a condition resulting in a heavy mass of the right-handed neutrino.
\section{Consistency of restrictions and Higgs boson masses}
\subsection{Minimization conditions and normalization conditions}
In the previous section, we analyzed only one minimization condition of Eq.~(\ref{1-4}). Usually, remaining minimization conditions Eqs.~(\ref{1-1}) $\sim$ (\ref{1-3}) are regarded as determining quadratic Higgs self-coupling constants $\mu_{1}$, $\mu_{2}$ and $\mu_{3}$.\cite{6} \ In our model, however, we must be more careful to use these conditions since the quadratic coupling constants are functions of the Yukawa coupling constants and the parameters introduced in ${\cal L}_{B}$. In addition to the minimization conditions, we also have five conditions, Eqs.~(\ref{1}) $\sim$ (\ref{5}), to normalize the kinetic terms of the gauge and Higgs fields. Furthermore, phenomenology of the Higgs boson masses will add several restrictions on the parameters of our model. In this subsection, we check the consistency among the parameter restrictions derived from the minimization conditions and normalization conditions.

Equations (\ref{1-1}) and (\ref{1-3}) of minimization conditions lead to complicated relations among many parameters containing $\eta_{a}^{l,q} (a=1\sim 4)$ and $C^{l,q}_{L,R~A,P}$. (See Eqs.~(\ref{1deriv-3}) and (\ref{1deriv-4}) in Appendix C.) We can use these to determine $\eta^{l}_{4}$ and $\eta^{q}_{4}$; there are no more restrictions on these two parameters. On the other hand, with Eq.~(\ref{phenomenological restriction}), Eq.~(\ref{1-2}) lead to
\begin{eqnarray}
\left|\frac{\epsilon_{1}\epsilon_{2}}{\epsilon_{+}^{2}}\frac{1}{tr(fh^{l}f^{\dag}\tilde{h}^{l})}\right.[&&\left.\frac{v_{L}^{2}+v_{R}^{2}}{\epsilon_{-}^{2}}tr\{f^{\dag}f(h^{l})^{2}-ff^{\dag}(\tilde{h}^{l})^{2}\}+tr\{(h^{l})^{4}+(\tilde{h^{l}})^{4}\}\right.\nonumber \\
&&\left.-\frac{(\alpha /\kappa )q_{1}+12q_{2}}{(\alpha /\kappa )l_{1}+12l_{2}}tr\{(h^{q})^{4}+(\tilde{h}^{q})^{4}\}]-\frac{h_{D1}^{2}}{2f_{1}^{2}}\right| < 10^{-8},\label{sub result}
\end{eqnarray}
where $l_{1}=\delta_{1}^{l}\bar{\delta }_{1}^{l}$, $l_{2}=\delta_{2}^{l}\bar{\delta }_{2}^{l}$, $q_{1}=\delta_{1}^{q}\bar{\delta }_{1}^{q}$ and $q_{2}=\delta_{2}^{q}\bar{\delta }_{2}^{q}$. Note that the left-hand sides of this equation, other than the term proportional to $tr\{f^{\dag}f(h^{l})^{2}-ff^{\dag}(\tilde{h}^{l})^{2}\}$, has a structure similar to the left-hand sides of Eq.~(\ref{main result}), namely $(h_{i}^{l,q})^{2}/(f_{i})^{2}$, roughly speaking.

Further, by taking several linear combinations of the normalization conditions Eqs.~(\ref{1}) $\sim$ (\ref{5}), we obtain the following independent restrictions:
\begin{eqnarray}
&&\frac{q_{1}+2(\alpha /\kappa )q_{2}}{l_{1}+2(\alpha /\kappa )l_{2}}=-\frac{9-36\,\mbox{sin}^{2}\theta _{W}}{9-20\,\mbox{sin}^{2}\theta _{W}},\label{restrict1}\\
&&\frac{l_{1}-6l_{2}}{q_{1}-6q_{2}}=\frac{tr\{(h^{q})^{2}+(\tilde{h}^{q})^{2}\}}{tr\{f^{\dag}f-(h^{l})^{2}-(\tilde{h}^{l})^{2}\}},\label{restrict2}\\
&&\kappa=-[12g_{2}^{2}\{l_{1}+q_{1}+2(\alpha/\kappa )(l_{2}+q_{2})\}]^{-1}\label{k},\\
&&\beta =\frac{4}{3}\{tr(f^{\dag}f)(l_{1}-6l_{2})\}^{-1},\\
&&(\frac{g_{3}}{g_{2}})^{2}=\frac{1}{4}\{1+\frac{l_{1}+2(\alpha /\kappa )l_{2}}{q_{1}+2(\alpha /\kappa )q_{2}}\},\label{3/2}
\end{eqnarray}
where $\theta_{W}$ is the Weinberg angle defined by $g_{1}^{2}/g_{2}^{2}=\mbox{sin}^{2}\theta_{W}/\mbox{cos}2\theta_{W}$ for the L-R symmetric model.\cite{7} \ We can use these restrictions to decrease the independent numbers of parameters as follows. The last three equations (\ref{k}) $\sim$ (\ref{3/2}) can be used to determine $\kappa$, $\beta$ and $(g_{3}/g_{2})^{2}$, respectively, none of which is contained in any other restrictions.\footnote{As independent parameters, we use $\alpha /\kappa $ and $\kappa$ instead of $\alpha$ and $\kappa$.} \ Moreover, Eqs.~(\ref{restrict1}) and (\ref{restrict2}) can be used to determine $\alpha /\kappa$ and $l_{2}$; we shall eliminate these two parameters from the other restrictions. Here we note that $(g_{3}/g_{2})^{2}$ can be determined by the Weinberg angle. Substituting Eq.~(\ref{restrict1}) into Eq.~(\ref{3/2}), we obtain
\begin{equation}
(\frac{g_{3}}{g_{2}})^{2}=-\frac{9-36\,\mbox{sin}^{2}\theta _{W}}{9-20\,\mbox{sin}^{2}\theta _{W}}\label{certain energy restriction},
\end{equation}
which implies $\mbox{sin}^{2}\theta_{W} > 0.25$ for a positive $(g_{3}/g_{2})^{2}$. We can adopt the point of view that parameter restrictions given by Sogami's method, such as Eqs.~(\ref{restrict1}) $\sim$ (\ref{3/2}), (\ref{zero}) and (\ref{mu 1}) $\sim$ (\ref{beta 1}), are tree level restrictions that hold at a certain energy scale $\mu_{0}$.\cite{9} \ Then, using the restrictions as initial conditions at $\mu_{0}$, we can study the evolution under the renormalization group of parameters. \ In this paper, however, we do not do the renormalization group analysis, and continue our discussion using the tree level restrictions together with the minimization conditions of the tree level Higgs potential.

Thus, remaining restrictions are Eqs.~(\ref{main result}) and (\ref{sub result}), under which we shall determine the independent parameters $l_{1}$, $q_{1}$ and $q_{2}$ and Yukawa coupling constants. Before we determine independent parameters and Yukawa coupling constants, let us decrease the independent numbers of Yukawa coupling constants using fermion mass expressions and their experimental values. For this purpose, in advance, we consider a condition of the Yukawa coupling constants ensuring the validity of Eqs.~(\ref{main result}) and (\ref{sub result}). Note that unless there are highly tuned cancellations on the left-hand sides of Eqs.~(\ref{main result}) and (\ref{sub result}), these equations may be satisfied in such a way: $f_{i}\gg h^{l,q}_{i}, \tilde{h}^{l,q}_{i}$ and $h^{l}_{i}-\tilde{h}^{l}_{i}\simeq 0$. Thus, in this subsection, we assume that the $h^{l}_{i}$ and $\tilde{h}^{l}_{i}$ are diagonal matrices satisfying
\begin{equation}
h^{l}_{i}=\tilde{h}^{l}_{i},\label{constraint-hl}
\end{equation}
under which the term proportional to $tr\{f^{\dag}f(h^{l})^{2}-ff^{\dag}(\tilde{h}^{l})^{2}\}$ of Eq.~(\ref{sub result}) vanishes. Further, from this point, we set
\begin{equation}
\frac{\epsilon_{1}}{\epsilon_{2}}=10.\label{tan b}
\end{equation}
Under Eqs.~(\ref{constraint-hl}) and (\ref{tan b}), we obtain
\begin{equation}
\frac{h_{3}^{q}}{\tilde{h}_{3}^{q}}=-\frac{339}{30} \mbox{~~or~} -\frac{401}{50}~,~~~ \left |\frac{h_{3}^{q}}{h_{3}^{l}}\right |\simeq  110\footnote{The resultant value of $\left |h_{3}^{q}/h_{3}^{l}\right |$ is sensitive to the choice of $h_{3}^{q}/\tilde{h}_{3}^{q}$. However, the above two values of $h_{3}^{q}/\tilde{h}_{3}^{q}$ lead to the same result in the approximation under consideration.}\label{110}
\end{equation}
for $m_{u^{3}}/m_{d^{3}}=40$ and $m_{e^{3}}/m_{d^{3}}=2/5$. We have used Eqs.~(\ref{e}) $\sim $ (\ref{d}) and $m_{W_{R}}\simeq 1.4\,\mbox{TeV}$ neglecting the generation mixings, and considered only to the third generations, since they are dominant in $h^{l,q}$ as $tr\{(h^{l})^{2}\}/(h_{3}^{l})^{2}-1<10^{-2}$. Owing to Eqs.~(\ref{constraint-hl}) and (\ref{110}), there is only one independent Yukawa coupling constant among $h^{l}_{3}, \tilde{h}^{l}_{3}, h_{3}^{q}$ and ${\tilde{h}_{3}^{q}}$; we choose $h^{l}_{3}$ as this constant.

Now, let us determine the parameter regions of $l_{1}$, $q_{1}$ and $q_{2}$ and the Yukawa coupling constants under Eqs.~(\ref{main result}) and (\ref{sub result}). First, we rewrite Eqs.~(\ref{main result}) and (\ref{sub result}) in terms of the independent parameters by using Eqs.~(\ref{restrict1}) and (\ref{restrict2}) in the forms of
\begin{eqnarray}
\frac{\alpha }{\kappa }&\simeq &-\frac{q_{1}+l_{1}/7}{2(q_{2}+l_{2}/7)},\label{a}\\
l_{2}&\simeq &\frac{1}{6}\{l_{1}-10^{5}\frac{(h^{l}_{3})^{2}}{|f_{3}|^{2}}\frac{q_{1}-6q_{2}}{8}\},\label{L2}
\end{eqnarray}
where we have used the numerical value $\mbox{sin}^{2}\theta_{W} \simeq 0.232$. By assuming the third generation dominance in $f$ and that each term on the left-hand sides of Eqs.~(\ref{main result}) and (\ref{sub result}) is of order $10^{-8}$, we obtain
\begin{eqnarray}
&&x<O(10^{-7})~,~~x\equiv \frac{(h^{l}_{3})^{2}}{|f_{3}|^{2}},\label{estimate1}\\
&&\left|\frac{q_{1}(q_{1}+l_{1}/7)-4q_{2}(6q_{2}+l_{1}/7)}{l_{1}(q_{1}-4q_{2}+l_{1}/21)}\right|<O(10^{-15}x^{-1})\label{order tune}
\end{eqnarray}
for a region of $6q_{2}-q_{1}\ll 10^{3}$. Equation (\ref{estimate1}) gives a lower bound of $m_{\nu_{R}^{3}}/m_{W_{R}}$ because of the relation
\begin{equation}
x\simeq 1.7\frac{m_{e^{3}}^{2}m_{W_{R}}^{2}}{m_{\nu_{R}^{3}}^{2}m_{W_{L}}^{2}},\label{N3}
\end{equation}
which is obtained from mass expressions of Eqs.~(\ref{e}) $\sim$ (\ref{WR mass}) and Eq.~(\ref{constraint-hl}). In particular, for $m_{W_{R}}\simeq 1.4\,\mbox{TeV}$, we obtain
\begin{equation}
m_{\nu_{R}^{3}}>10^{5}\,\mbox{GeV}
\end{equation}
by using $m_{e^{3}}/m_{W_L} \simeq 2.3\times 10^{-2}$. A small $x$ also leads to a large upper bound of the left-hand side of Eq.~(\ref{order tune}), which allows large regions of $l_{1}$, $q_{1}$ and $q_{2}$. If we set $m_{\nu_{R}^{3}}$ to a mass of the intermediate scale in GUT-breaking scenarios,\cite{8} \ e.\,g., $10^{13}\,\mbox{GeV}$, we will obtain $O(x)\sim 10^{-23}$ from Eq.~(\ref{N3}). Then Eq.~(\ref{order tune}) requires no fine-tuning.
\subsection{Higgs boson masses}
Under the minimization conditions and normalization conditions, we were able to determine the parameters of our model without fine-tuning by requiring the right-handed tau neutrino mass to be sufficiently heavy. In this subsection, we give expressions of the Higgs boson masses, and in the next subsection, we consider tuning of parameters taking phenomenology of the Higgs boson masses into account.

To obtain expressions for the Higgs boson masses, we use the general forms of mass square matrices $M^{Re}$, $M^{Im}$, $M^{+}$ and $M^{++}$ given in Appendix of Ref.~\citen{6}, which correspond to the bases $\{\phi_{1}^{0r}, \phi_{2}^{0r}, \delta_{R}^{0r}, \delta_{L}^{0r} \}$, $\{\phi_{1}^{0i}, \phi_{2}^{0i}, \delta_{R}^{0i}, \delta_{L}^{0i} \}$, $\{\phi_{1}^{+}, \phi_{2}^{+}, \delta_{R}^{+}, \delta_{L}^{+} \}$ and $\{\delta_{R}^{++}, \delta_{L}^{++} \}$, respectively.\footnote{We have used definitions such as $\phi_{1}^{0}=(\phi_{1}^{0r}+i\phi_{1}^{0i})/\sqrt{2}$.} \ Under the conditions of Eq.~(\ref{tuning-1}), respective eigenvalues of the mass matrices become\footnote{We derived these results with the aid of MATHEMATICA.}
\begin{eqnarray}
M^{Re}&~:~& m_{H}^{2}, ~~m_{Re1}^{2}, ~~{\tilde m}_{H}^{2}, ~~m_{Re2}^{2},\\
M^{Im}&~:~& m_{H}^{2},  ~~0, ~~0, ~~0,\\
M^{+}&~:~& m_{H}^{2}, ~~m_{+}^{2}, ~~0, ~~0,\\
M^{++}&~:~& m_{H}^{2},~~m_{H}^{2},
\end{eqnarray}
where
\begin{eqnarray}
m_{H}^{2}&=&\frac{3}{8}(\alpha l_{1}+12\kappa l_{2})v_{R}^{2}tr\{(f^{\dag}f)^{2}\},\label{MH}\\
m_{Re1}^{2}&=&\frac{3}{2}(\alpha l_{1}+12\kappa l_{2})\frac{\epsilon_{1}^{2}\epsilon_{2}^{2}}{\epsilon_{+}^{2}}\frac{v_{R}^{2}}{\epsilon_{-}^{2}}tr\{(h^{l})^{2}f^{\dag}f-({\tilde h^{l}})^{2}ff^{\dag}\},\label{MR1}\\
m_{+}^{2}&=&\frac{3}{8}(\alpha l_{l}+12\kappa l_{2})\epsilon_{+}^{2}\frac{v_{R}^{2}}{\epsilon_{-}^{2}}tr\{(h^{l})^{2}f^{\dag}f-({\tilde h^{l}})^{2}ff^{\dag}\},\label{M+}
\\
{\tilde m}_{H}^{2}&\simeq &2v_{R}^{2}(|f_{3}|^{2})^{2}\{\eta_{1}^{l}\eta_{2}^{l}-\frac{3}{8}(\alpha l_{1}+12\kappa l_{2})\},\label{MHT}\\
m_{Re2}^{2}&\simeq &2\epsilon_{+}^{2}(h_{3}^{q})^{4}\{\eta_{1}^{q}\eta_{2}^{q}-\frac{3}{8}(\alpha q_{1}+12\kappa q_{2})-\frac{\dfrac{1}{4}(\eta_{3}^{l}\eta_{3}^{q})^{2}}{\eta_{1}^{l}\eta_{2}^{l}-\dfrac{3}{8}(\alpha l_{1}+12\kappa l_{2})}\}.~~\label{MR2}
\end{eqnarray}
These expressions are only the leading terms obtained under the third generation dominance in the Yukawa coupling matrices and several conditions mentioned below.

First, we have used Eq~(\ref{seesaw}). In particular, we have set $v_{L}$ to zero in $M^{Re}$ before calculation of its eigenvalues. We have also dropped a term proportional to $v_{L}$ in the expression of $m_{+}^{2}$. Second, we have used the inequalities
\begin{equation}
\alpha_{3}\ll \lambda_{1}\ll \alpha_{1}\ll \rho_{1}, \rho_{2}\label{inequalities}.
\end{equation}
These inequalities are understood in terms of the concrete expressions of each parameter given in Appendix B. Under the condition of Eq.~(\ref{tuning-1}), they reduce to
\begin{eqnarray}
\alpha_{3}&=&\frac{3}{4}(\alpha l_{1}+12\kappa l_{2})tr\{(h^{l})^{2}f^{\dag}f-({\tilde h})^{2}ff^{\dag}\},\\
\lambda_{1}&\simeq &-\{\frac{3}{8}(\alpha q_{1}+12\kappa q_{2})-\eta_{1}^{q}\eta_{2}^{q}\}(h^{q}_{3})^{4}\nonumber \\
&&-\{\frac{3}{4}(\alpha l_{1}+12\kappa l_{2})-4\eta_{1}^{l}\eta_{2}^{l}\}(h^{l}_{3})^{4}-2\eta_{3}^{l}\eta_{3}^{q}(h^{l}_{3}h^{q}_{3})^{2},\\
\alpha_{1}&\simeq &|f_{3}|^{2} [\eta_{3}^{l}\eta_{3}^{q}(h^{q}_{3})^{2}-\{\frac{3}{4}(\alpha l_{1}+12\kappa l_{2})-4\eta_{1}^{l}\eta_{2}^{l}\}(h^{l}_{3})^{2}],\\
\rho_{1}&\simeq &-(|f_{3}|^{2})^{2}\{\frac{3}{8}(\alpha l_{1}+12\kappa l_{2})-\eta_{1}^{l}\eta_{2}^{l}\},\\
\rho_{2}&=&\frac{3}{16}(\alpha l_{1}+12\kappa l_{2})tr\{(f^{\dag}f)^{2}\}.
\end{eqnarray}
If we choose appropriate values for $\alpha q_{1}+12\kappa q_{2}$ and undetermined parameters $\eta_{a}^{l,q}~(a=1\sim 3)$, these expressions explain the inequalities (\ref{inequalities}), except the reason for the fact that $\alpha_{3}$ is the smallest parameter, due to Eqs.~(\ref{110}) and (\ref{estimate1}). The maximally small nature of $\alpha_{3}$, as expressed in the inequalities (\ref{inequalities}) is derived from a condition that makes Eq.~(\ref{sub result}) hold. The condition of Eq.~(\ref{constraint-hl}) leads to $\alpha_{3}=0$. Since the vanishing of $\alpha_{3}$ gives rise to the unpleasant result $m_{Re1}^{2}=m_{+}^{2}=0$, we modify Eq.~(\ref{constraint-hl}) to
\begin{equation}
\frac{v_{R}^{2}}{\epsilon_{-}^{2}}tr\{(h^{l})^{2}f^{\dag}f-({\tilde h^{l}})^{2}ff^{\dag}\}\simeq tr\{(h^{l})^{4}+(\tilde{h^{l}})^{4}\}\label{nonzero}
\end{equation}
to obtain non-zero $\alpha_{3}$ with the inequalities (\ref{inequalities}). In this case, Eqs.~(\ref{main result}) and (\ref{sub result}) yield $({\tilde h^{l}_{3}})^{2}/(h^{l}_{3})^{2}-1 \sim O(10^{-9})$ together with Eq.~(\ref{estimate1}). The small difference between $h^{l}$ and ${\tilde h}^{l}$ does not affect the analyses that we have done using Eq.~(\ref{constraint-hl}) in the previous subsection. Finally, to obtain the expression of Eq.~(\ref{MR2}) from the result of calculation of $m_{Re2}^{2}$, we have chosen $\eta_{a}^{l,q}~(a=1\sim 3)$ so that Eq.~(\ref{110}) retains dominance of the term proportional to $(h_3^q)^4$ in the expression of $m_{Re2}^{2}$.
\subsection{Higgs boson masses and parameter tuning}
We can reduce each mass expression of Eqs.~(\ref{MH}) $\sim$ (\ref{MR2}) to
\begin{eqnarray}
m_{Re1}^{2}&\simeq &(m_{+}/5)^{2}\simeq  3.3\times 10^{-1}\frac{m_{e^{3}}^{4}}{m_{W_{L}}^{2}}g_{2}^{2}(\alpha l_{1}+12\kappa l_{2}),\label{MR1 and M+} \\
m_{H}^{2}&\simeq &\frac{3}{4}\frac{m_{\nu_{R}^{3}}^{4}}{m_{W_{R}}^{2}}g_{2}^{2}(\alpha l_{1}+12\kappa l_{2}),\label{MH2} \\
{\tilde m}_{H}^{2}&\simeq &4\frac{m_{\nu_{R}^{3}}^{4}}{m_{W_{R}}^{2}}g_{2}^{2}\{\eta_{1}^{l}\eta_{2}^{l}-\frac{3}{8}(\alpha l_{1}+12\kappa l_{2})\},\label{MH2-2}\\
m_{Re2}^{2}&\simeq &3.4\times 10^{1}\frac{m_{u^{3}}^{4}}{m_{W_{L}}^{2}}g_{2}^{2}\{\eta_{1}^{q}\eta_{2}^{q}-\frac{3}{8}(\alpha q_{1}+12\kappa q_{2})\nonumber \\
&&~~~~~~~~~~~~~~~~~~~~~~~-\frac{\frac{1}{4}(\eta_{3}^{l}\eta_{3}^{q})^{2}}{\eta_{1}^{l}\eta_{2}^{l}-\frac{3}{8}(\alpha l_{1}+12\kappa l_{2})}\},\label{MR2-2}
\end{eqnarray}
by using Eqs.~(\ref{tan b}), (\ref{110}),\footnote{Whichever of the values we choose for ${\tilde h}_{3}^{q}/h_{3}^{q}$, we obtain the same result in the approximation under consideration.} \ (\ref{nonzero}) and (\ref{e}) $\sim$ (\ref{WR mass}). With the phenomenology of the Higgs boson masses given in the following, these expressions lead to several restrictions on the independent parameters $l_{1}$, $q_{1}$, $q_{2}$, $x$ and $\eta_{a}^{l,q}~(a=1\sim 3)$.

First, $m_{Re1}\gsim 10\,\mbox{TeV}$ is necessary for the suppression of the effect of flavor-changing neutral currents (FCNC) of quarks since the eigenstate of $m_{Re1}^{2}$, $(\epsilon_{1}\phi_{2}^{0r}-\epsilon_{2}\phi_{1}^{0r})/\epsilon_{+}$, can couple to them.\cite{6,FCNC} \ Therefore, with $m_{e^{3}}^{4}/m_{W_{L}}^{2}\simeq 1.6 \times 10^{-3}(\mbox{GeV})^{2}$, Eq.~(\ref{MR1 and M+}) leads to
\begin{equation}
g_{2}^{2}(\alpha l_{1}+12\kappa l_{2})\gsim 1.9\times 10^{11}.\label{Higgs restriction}
\end{equation}
Substituting Eqs.~(\ref{k}), (\ref{a}) and (\ref{L2}) into this inequality, we obtain the new restriction
\begin{equation}
\frac{q_{1}+l_{1}/7-2(6q_{2}-q_{1})}{6q_{2}-q_{1}}\gsim 2.0\times 10^{12}.\label{tuning a}
\end{equation}

Second, by setting the upper bound of $m_{H}$ to $10^{19}\,\mbox{GeV}$ and using $m_{W_{R}}\simeq 1.4\mbox{TeV}$, we obtain
\begin{equation}
m_{\nu_{R}^{3}}\lsim 1.9\times 10^{8}\mbox{GeV}
\end{equation}
from Eqs.~(\ref{MH2}) and (\ref{Higgs restriction}). Owing to Eq.~(\ref{N3}), this inequality gives
\begin{equation}
x\gsim  4.7\times 10^{-14}.\label{x}
\end{equation}

Third, from Eqs.~(\ref{MH2-2}) and (\ref{MR2-2}), we can obtain two restrictions on $\eta_{a}^{l,q}~(a=1\sim 3)$. We shall consider these restrictions after consideration of tuning of $l_{1}$, $q_{1}$, $q_{2}$ and $x$.

Now, let us consider the tuning of $l_{1}$, $q_{1}$, $q_{2}$ and $x$ under all the restrictions, Eqs.~(\ref{estimate1}), (\ref{order tune}), (\ref{tuning a}) and (\ref{x}). Equations (\ref{estimate1}) and (\ref{x}) yield the following range of $x$:
\begin{equation}
O(10^{-7})>x>O(10^{-14}),\label{range of x}
\end{equation}
which corresponds to 
\begin{equation}
O(10^{5})\,\mbox{GeV}<m_{\nu_{R}^{3}}<O(10^{8})\,\mbox{GeV}.
\end{equation}
Equations (\ref{order tune}), (\ref{tuning a}) and (\ref{x}) require highly tuned cancellations or hierarchical tuning of the parameters. For example, if we put $x \sim O(10^{-13})$, we can have
 \begin{eqnarray}
&&l_{1}\sim O(1),\label{l_1}\\
&&q_{1}\sim O(10^{-2}),\label{q_l}\\
&&0<6q_{2}-q_{1}\lsim O(10^{-13}).\label{fine-tuning}
\end{eqnarray}
The small difference between $6q_{2}$ and $q_{1}$ is important since it determines the mass of the FCNC coupling Higgs boson $m_{Re1}$ ($6q_{2}-q_{1} \to 0$ corresponds to $m_{Re1} \to \infty $.); such a choice of $q_{1}$ and $q_{2}$ is fine-tuning.

Finally, let us tune $\eta_{a}^{l,q}$ $(a=1\sim 3)$ in Eqs.~(\ref{MH2-2}) and (\ref{MR2-2}). We can show that each eigenstate of ${\tilde m}_{H}^{2}$ and $m_{Re2}^{2}$ is a combination of $(\epsilon_{1}\phi_{1}^{0r}+\epsilon_{2}\phi_{2}^{0r})/\epsilon_{+}$ and $\delta_{R}^{0r}$. The state $(\epsilon_{1}\phi_{1}^{0r}+\epsilon_{2}\phi_{2}^{0r})/\epsilon_{+}$ is an analogue of the standard model Higgs boson state\cite{6} and is dominant over the eigenstate of $m_{Re2}^{2}$. Thus, we should set
\begin{equation}
m_{Re2}\sim O(10^{2}\,\mbox{GeV}).\label{standard Higgs mass}
\end{equation}
On the other hand, by using Eqs.~(\ref{k}), (\ref{a}), (\ref{L2}) and (\ref{l_1}) $\sim$ (\ref{fine-tuning}), we obtain
\begin{eqnarray}
g_{2}^{2}(\alpha q_{1}+12\kappa q_{2})\simeq \frac{7q_{1}(q_{1}+l_{1}/7)}{72l_{1}(6q_{2}-q_{1})}>O(10^{9}).\label{qq}
\end{eqnarray}
Equations (\ref{standard Higgs mass}) and (\ref{qq}) require a cancellation on the right-hand sides of Eq.~(\ref{MR2-2}). As a possible tuning, we set
\begin{eqnarray}
\eta_{1}^{l}\eta_{2}^{l}&\equiv &\frac{9}{16}(\alpha l_{1}+12\kappa l_{2}),\label{eta1}\\
\eta_{1}^{q}\eta_{2}^{q}&\equiv &\frac{9}{16}(\alpha q_{1}+12\kappa q_{2}),\\
\eta_{3}^{l}\eta_{3}^{q}&\equiv &\frac{3}{8}\{(\alpha l_{1}+12\kappa l_{2})(\alpha q_{1}+12\kappa q_{2})(1-\xi)\}^{\frac{1}{2}}\label{eta3},
\end{eqnarray}
where $\xi$ is a parameter determined.\footnote{If respective values of $g_{2}^{2}(\alpha l_{1}+12\kappa l_{2})$, $g_{2}^{2}(\alpha q_{1}+12\kappa q_{2})$ and $\eta_{a}^{l,q} (a=1\sim 3)$ are of the same order, inequalities (\ref{inequalities}) hold, and the expression of Eq.~(\ref{MR2}) is valid. Equations (\ref{Higgs restriction}), (\ref{qq}), (\ref{eta1}) $\sim$ (\ref{eta3}) allow for such values.} \ In this case, Eqs.~(\ref{MH2-2}) and (\ref{MR2-2}) become
\begin{eqnarray}
{\tilde m}_{H}^{2}&\simeq  &m_{H}^{2},\\
m_{Re2}^{2}&\simeq &6.4\frac{m_{u^{3}}^{4}}{m_{W_{L}}^{2}}\xi g_{2}^{2}(\alpha q_{1}+12\kappa q_{2})\label{Re2},
\end{eqnarray}
respectively. Substituting Eqs.~(\ref{standard Higgs mass}) and (\ref{qq}) into Eq.~(\ref{Re2}) and using $m_{u^{3}}^{4}/m_{W_{L}}^{2}\\ \simeq 1.5\times10^{5}\,(\mbox{GeV})^{2}$, we obtain
\begin{equation}
0<\xi \lsim O(10^{-11}).\label{m}
\end{equation}
This implies the necessity of fine-tuning of $\eta_{a}^{l,q} (a=1\sim 3)$.

Now, two zero eigenvalue states of $M^{+}$ and $M^{Im}$ correspond to Nambu-Goldstone bosons coming from gauge symmetry breaking, respectively. \cite{6} \ However, one more zero eigenvalue state of $M^{Im}$ belongs to the physical Higgs bosons. Exclusion of such a massless Higgs boson and the fine-tuning such as Eqs.~(\ref{l_1}) $\sim$ (\ref{fine-tuning}) and (\ref{m}) are remaining problems, which should be solved in the future.
\section{Conclusion}
We have studied an application of Sogami's generalized covariant derivative method to a ${SU(2)}_L\times{SU(2)}_R\times{U(1)}_{B-L}\times SU(3)_{c}$ model, and the minimization condition of the Higgs potential is discussed in detail within the framework of classical theory.

We have shown that a restriction, which has been known as the restriction requiring fine-tuning of Higgs-self-coupling constants, can be reduced to a condition of Yukawa coupling constants giving a heavy mass of the right-handed tau neutrino. We have also discussed the consistency among the parameter restriction derived from the minimization conditions and the normalization conditions for kinetic terms. Then by setting the mass of the right-handed tau neutrino to be sufficiently heavy, we could determine those parameters without fine-tuning.

However, we have encountered problems: First, if we attempt to assign the order of $\gsim 10\,\mbox{TeV}$ and $10^2\,\mbox{GeV}$, respectively, to the masses of a FCNC Higgs boson and the analogue of the standard model Higgs boson, we need fine-tuning. Second, after the tuning of parameters, there still remains a massless neutral Higgs boson. This massless Higgs boson can be understood in the following sense: Except for the Yukawa interaction term ${\cal L}_{Y}$, our lagrangian has a global $U(1)$ symmetry such as ${\bf \Phi} \to e^{i\theta}{\bf \Phi}$. A breaking of this symmetry by the non-zero VEV of ${\bf \Phi}$ leads to a NG boson in the ${\cal L}_{Y}$-excepted system. Since the presence of ${\cal L}_{Y}$ does not affect the Higgs boson masses at the tree level, such a zero mass state will also appear in the full lagrangian system.

We note that if we consider our lagrangian to be the bare lagrangian, the renomalizability of our model is not trivial, since our Higgs potential lacks several terms allowed by the symmetry and some coupling constants are not independent of each other. As mentioned in \S 4, we can adopt the point of view that the parameter restrictions of our model are the tree level restrictions that hold at a certain energy scale.\cite{9} \ Then we can study a renormalization group analysis of parameters of our model. The above problems should be discussed in a future work with this analysis.
\section*{Acknowledgements}
The author wishes to express his sincere thanks to Professor S.~Naka for discussions and careful reading of the manuscript. He is grateful to Professor S.~Ishida, Professor J.~Otokozawa, Professor S.~Y.~Tsai, Dr.~S.~Deguchi, Dr.~S.~Kamefuchi, Dr.~M.~Sekiguchi and other members of his laboratory for their encouragement and useful comments. He also thanks the late Professor O.~Hara for his interest and encouragement.
\appendix
\section{Generalized Field Strengths}
We give generalized field strengths of our model. With decomposition such as
\begin{eqnarray}
\check{F}_{\mu \nu}^{l}&=&-i\sum_{k}^{1,2}g_{k}F_{\mu \nu}^{l(k)}+\frac{i}{4}F_{\mu \nu}^{l(0)},\\
\check{F}_{\mu \nu}^{q}&=&-i\sum_{k}^{1,2,3}g_{k}F_{\mu \nu}^{q(k)}+\frac{i}{4}F_{\mu \nu}^{q(0)},
\end{eqnarray}
we obtain
\begin{eqnarray}
F_{\mu \nu}^{l(1)}&=&-\frac{1}{2}\mbox{diag}(F_{\mu \nu}^{B}{\bf 1},-F_{\mu \nu}^{B}{\bf 1},F_{\mu \nu}^{B}{\bf 1},-F_{\mu \nu}^{B}{\bf 1})\otimes 1_{g}\otimes 1_{c},\\
F_{\mu \nu}^{q(1)}&=&\frac{1}{6}\mbox{diag}(F_{\mu \nu}^{B}{\bf 1},-F_{\mu \nu}^{B}{\bf 1},F_{\mu \nu}^{B}{\bf 1},-F_{\mu \nu}^{B}{\bf 1})\otimes 1_{g}\otimes 1_{c},\\
F_{\mu \nu}^{l(2)}&=&F_{\mu \nu}^{q(2)}=\mbox{diag}({\bf F}_{\mu \nu}^{L},-{\bf F}_{\mu \nu}^{L \ast},{\bf F}_{\mu \nu}^{R},{-\bf F}_{\mu \nu}^{R \ast})\otimes 1_{g}\otimes 1_{c},\\
F_{\mu \nu}^{q(3)}&=&\mbox{diag}(G_{\mu \nu}{\bf 1},-G_{\mu \nu}^{\ast}{\bf 1},G_{\mu \nu}{\bf 1},-G_{\mu \nu}^{\ast}{\bf 1})\otimes 1_{g}\otimes 1_{c},\\
F_{\mu \nu}^{l,q(0)}&=&\gamma_{\nu}D_{\mu}A^{l,q(0)}-\gamma_{\mu}D_{\nu}A^{l,q(0)}+\frac{1}{2}\sigma_{\mu \nu}(A^{l,q(0)})^{2},
\end{eqnarray}
where
\begin{eqnarray*}
D_{\mu}A^{l(0)}&=&{\cal E}\pmatrix{0&f^{\dag}(D_{\mu}{\bf \Delta}_L)^{\dag}&h^{l}D_{\mu}{\bf \Phi}&0\cr fD_{\mu}{\bf \Delta}_L &0&0&{\tilde h}^{l}D_{\mu}{\bf \Phi}\cr h^{l}(D_{\mu}{\bf \Phi})^{\dag}&0&0&f^{\dag}(D_{\mu}{\bf \Delta}_R)^{\dag}\cr 0&{\tilde h}^{l}(D_{\mu}{\bf \Phi})^{\dag}&fD_{\mu}{\bf \Delta}_R &0\cr}{\cal E}^{\dag},\\
D_{\mu}A^{q(0)}&=&{\cal E}\pmatrix{0&0&h^{q}D_{\mu}{\bf \Phi}&0\cr 0&0&0&{\tilde h}^{q}D_{\mu}{\bf \Phi}\cr h^{q}(D_{\mu}{\bf \Phi})^{\dag}&0&0&0\cr 0&{\tilde h}^{q}(D_{\mu}{\bf \Phi})^{\dag}&0&0\cr }{\cal E}^{\dag},\\
(A^{l,q(0)})^{2}&=&{\cal E}\pmatrix{(A^{l,q(0)})^{2}_{LL}&(A^{l,q(0)})^{2}_{LR}\cr (A^{l,q(0)})^{2}_{RL}&(A^{l,q(0)})^{2}_{RR}\cr}{\cal E}^{\dag}
\end{eqnarray*}
and
\begin{eqnarray*}
(A^{l(0)})^{2}_{LL}&=&\pmatrix{f^{\dag}f{\bf \Delta}_{L}^{\dag}{\bf \Delta}_{L}+(h^{l})^{2}{\bf \Phi}{\bf \Phi}^{\dag}+(C_{LP}^{l})^{2}{\bf 1}~,~(C_{LP}^{l}f^{\dag}+f^{\dag}C_{LA}^{l}){\bf \Delta}_{L}^{\dag}\cr (fC_{LP}^{l}+C_{LA}^{l}f){\bf \Delta}_{L}~,~ff^{\dag}{\bf \Delta}_{L}{\bf \Delta}_{L}^{\dag}+({\tilde h}^{l})^{2}{\bf \Phi}{\bf \Phi}^{\dag}+(C_{LA}^{l})^{2}{\bf 1}\cr},\\
(A^{l(0)})^{2}_{RR}&=&\pmatrix{f^{\dag}f{\bf \Delta}_{R}^{\dag}{\bf \Delta}_{R}+(h^{l})^{2}{\bf \Phi}^{\dag}{\bf \Phi}+(C_{RP}^{l})^{2}{\bf 1}~,~(C_{RP}^{l}f^{\dag}+f^{\dag}C_{RA}^{l}){\bf \Delta}_{R}^{\dag}\cr (fC_{RP}^{l}+C_{RA}^{l}f){\bf \Delta}_{R}~,~ff^{\dag}{\bf \Delta}_{R}{\bf \Delta}_{R}^{\dag}+({\tilde h}^{l})^{2}{\bf \Phi}^{\dag}{\bf \Phi}+(C_{RA}^{l})^{2}{\bf 1}\cr},\\
(A^{l(0)})^{2}_{LR}&=&\pmatrix{(C_{LP}^{l}h^{l}+h^{l}C_{RP}^{l}){\bf \Phi}&f^{\dag}{\tilde h}^{l}{\bf \Delta}_{L}^{\dag}{\bf \Phi}+h^{l}f^{\dag}{\bf \Phi}{\bf \Delta}^{\dag}_{R}\cr fh^{l}{\bf \Delta}_{L}{\bf \Phi}+{\tilde h}^{l}f{\bf \Phi}{\bf \Delta}_{R}&(C_{LA}^{l}{\tilde h}^{l}+{\tilde h}^{l}C_{RA}^{l}){\bf \Phi}\cr},\\
(A^{l(0)})^{2}_{RL}&=&\pmatrix{(h^{l}C_{LP}^{l}+C_{RP}^{l}h^{l}){\bf \Phi}^{\dag}&fh^{l}{\bf \Delta}_{L}^{\dag}{\bf \Phi}+h^{l}f^{\dag}{\bf \Phi}{\bf \Delta}^{\dag}_{R}\cr (f^{\dag}{\tilde h}^{l}{\bf \Delta}_{L}^{\dag}{\bf \Phi}+h^{l}f^{\dag}{\bf \Phi}{\bf \Delta}^{\dag}_{R})^{\dag}&({\tilde h}^{l}C_{LA}^{l}+C_{RA}^{l}{\tilde h}^{l}){\bf \Phi}^{\dag}\cr},\\
(A^{q(0)})^{2}_{LL}&=&\pmatrix{(h^{q})^{2}{\bf \Phi}{\bf \Phi}^{\dag}+(C_{LP}^{q})^{2}{\bf 1}&0\cr 0&({\tilde h}^{q})^{2}{\bf \Phi}{\bf \Phi}^{\dag}+(C_{LA}^{q})^{2}{\bf 1}\cr},\\
(A^{q(0)})^{2}_{RR}&=&\pmatrix{(h^{q})^{2}{\bf \Phi}^{\dag}{\bf \Phi}+(C_{RP}^{q})^{2}{\bf 1}&0\cr 0&({\tilde h}^{q})^{2}{\bf \Phi}^{\dag}{\bf \Phi}+(C_{RA}^{q})^{2}{\bf 1}\cr},\\
(A^{q(0)})^{2}_{LR}&=&\pmatrix{(C_{LP}^{q}h^{q}+h^{q}C_{RP}^{q}){\bf \Phi}&0\cr 0&(C_{LA}^{q}{\tilde h}^{q}+{\tilde h}^{q}C_{RA}^{q}){\bf \Phi}\cr},\\
(A^{q(0)})^{2}_{RL}&=&\pmatrix{(h^{q}C_{LP}^{q}+C_{RP}^{q}h^{q}){\bf \Phi}^{\dag}&0\cr 0&({\tilde h}^{q}C_{LA}^{q}+C_{RA}^{q}{\tilde h}^{q}){\bf \Phi}^{\dag}\cr }.
\end{eqnarray*}
\section{Higgs Self-Coupling Constants}
Here we give the non-vanishing Higgs self-coupling constants of our model.
\begin{eqnarray}
-\mu_{1}^{2}&=&-\frac{3}{16}\alpha tr[2\{(C_{P}^{l})^{2}\bar{\delta}_{P1}^{l}(h^{l})^{2}\delta_{P1}^{l}+(C_{P}^{l})^{2}\delta_{P1}^{l}(h^{l})^{2}\bar{\delta}_{P1}^{l}\nonumber \\
&&~~~~~~~~~~~~~\,~+(C_{A}^{l})^{2}\bar{\delta}_{A1}^{l}(\tilde{h}^{l})^{2}\delta_{A1}^{l}+(C_{A}^{l})^{2}\delta_{A1}^{l}(\tilde{h}^{l})^{2}\bar{\delta}_{A1}^{l}\}\nonumber \\
&&~~~~~~~~~~~+2\{(C_{P}^{l}h^{l}+h^{l}C_{P}^{l})\bar{\delta}_{P1}^{l}(C_{P}^{l}h^{l}+h^{l}C_{P}^{l})\delta_{P1}^{l}\nonumber \\
&&~~~~~~~~~~~~~~~\,~+(C_{A}^{l}{\tilde h}^{l}+{\tilde h}^{l}C_{A}^{l})\bar{\delta}_{A1}^{l}(C_{A}^{l}{\tilde h}^{l}+{\tilde h}^{l}C_{A}^{l})\delta_{A1}^{l}\}\nonumber \\
&&~~~~~~~~~~~+\{(h^{q})^{2}\bar{\delta}_{P1}^{q}(C_{LP}^{q})^{2}\delta_{P1}^{q}+
(C_{LP}^{q})^{2}\bar{\delta}_{P1}^{q}(h^{q})^{2}\delta_{P1}^{q}\nonumber \\
&&~~~~~~~~~~~~~~~+(\tilde{h}^{q})^{2}\bar{\delta}_{A1}^{q}(C_{LA}^{q})^{2}\delta_{A1}^{q}+
(C_{LA}^{q})^{2}\bar{\delta}_{A1}^{q}(\tilde{h}^{q})^{2}\delta_{A1}^{q}\}\nonumber \\
&&~~~~~~~~~~~+\{(h^{q})^{2}\bar{\delta}_{P1}^{q}(C_{RP}^{q})^{2}\delta_{P1}^{q}+
(C_{RP}^{q})^{2}\bar{\delta}_{P1}^{q}(h^{q})^{2}\delta_{P1}^{q}\nonumber \\
&&~~~~~~~~~~~~~~~+(\tilde{h}^{q})^{2}\bar{\delta}_{A1}^{q}(C_{RA}^{q})^{2}\delta_{A1}^{q}+
(C_{RA}^{q})^{2}\bar{\delta}_{A1}^{q}(\tilde{h}^{q})^{2}\delta_{A1}^{q}\}\nonumber \\
&&~~~~~~~~~~~+\{(C_{LP}^{q}h^{q}+h^{q}C_{RP}^{q})\bar{\delta}_{P1}^{q}(C_{RP}^{q}h^{q}+h^{q}C_{LP}^{q})\delta_{P1}^{q}\nonumber \\
&&~~~~~~~~~~~~~~~+(C_{LA}^{q}{\tilde h}^{q}+{\tilde h}^{q}C_{RA}^{q})\bar{\delta}_{A1}^{q}(C_{RA}^{q}{\tilde h}^{q}+{\tilde h}^{q}C_{LA}^{q})\delta_{A1}^{q}\}\nonumber \\
&&~~~~~~~~~~~+\{(C_{LP}^{q}h^{q}+h^{q}C_{RP}^{q})\delta_{P1}^{q}(C_{RP}^{q}h^{q}+h^{q}C_{LP}^{q})\bar{\delta}_{P1}^{q}\nonumber \\
&&~~~~~~~~~~~~~~~+(C_{LA}^{q}{\tilde h}^{q}+{\tilde h}^{q}C_{RA}^{q})\delta_{A1}^{q}(C_{RA}^{q}{\tilde h}^{q}+{\tilde h}^{q}C_{LA}^{q})\bar{\delta}_{A1}^{q}\}]\nonumber \\
&&-\frac{9}{4}\kappa tr[2\{(C_{P}^{l})^{2}\bar{\delta}_{P2}^{l}(h^{l})^{2}\delta_{P2}^{l}+(C_{P}^{l})^{2}\delta_{P2}^{l}(h^{l})^{2}\bar{\delta}_{P2}^{l}\nonumber \\
&&~~~~~~~~~~~~~+(C_{A}^{l})^{2}\bar{\delta}_{A2}^{l}(\tilde{h}^{l})^{2}\delta_{A2}^{l}+(C_{A}^{l})^{2}\delta_{A2}^{l}(\tilde{h}^{l})^{2}\bar{\delta}_{A2}^{l}\}\nonumber \\
&&~~~~~~~~\,~+2\{(C_{P}^{l}h^{l}+h^{l}C_{P}^{l})\bar{\delta}_{P2}^{l}(C_{P}^{l}h^{l}+h^{l}C_{P}^{l})\delta_{P2}^{l}\nonumber \\
&&~~~~~~~~~~~~~~~+(C_{A}^{l}{\tilde h}^{l}+{\tilde h}^{l}C_{A}^{l})\bar{\delta}_{A2}^{l}(C_{A}^{l}{\tilde h}^{l}+{\tilde h}^{l}C_{A}^{l})\delta_{A2}^{l}\}\nonumber \\
&&~~~~~~~~\,~+\{(h^{q})^{2}\bar{\delta}_{P2}^{q}(C_{LP}^{q})^{2}\delta_{P2}^{q}+
(C_{LP}^{q})^{2}\bar{\delta}_{P2}^{q}(h^{q})^{2}\delta_{P2}^{q}\nonumber \\
&&~~~~~~~~~~~~\,~+(\tilde{h}^{q})^{2}\bar{\delta}_{A2}^{q}(C_{LA}^{q})^{2}\delta_{A2}^{q}+
(C_{LA}^{q})^{2}\bar{\delta}_{A2}^{q}(\tilde{h}^{q})^{2}\delta_{A2}^{q}\}\nonumber \\
&&~~~~~~~~\,~+\{(h^{q})^{2}\bar{\delta}_{P2}^{q}(C_{RP}^{q})^{2}\delta_{P2}^{q}+
(C_{RP}^{q})^{2}\bar{\delta}_{P2}^{q}(h^{q})^{2}\delta_{P2}^{q}\nonumber \\
&&~~~~~~~~~~~~\,~+(\tilde{h}^{q})^{2}\bar{\delta}_{A2}^{q}(C_{RA}^{q})^{2}\delta_{A2}^{q}+
(C_{RA}^{q})^{2}\bar{\delta}_{A2}^{q}(\tilde{h}^{q})^{2}\delta_{A2}^{q}\}\nonumber \\
&&~~~~~~~~\,~+\{(C_{LP}^{q}h^{q}+h^{q}C_{RP}^{q})\bar{\delta}_{P2}^{q}(C_{RP}^{q}h^{q}+h^{q}C_{LP}^{q})\delta_{P2}^{q}\nonumber \\
&&~~~~~~~~~~~~\,~+(C_{LA}^{q}{\tilde h}^{q}+{\tilde h}^{q}C_{RA}^{q})\bar{\delta}_{A2}^{q}(C_{RA}^{q}{\tilde h}^{q}+{\tilde h}^{q}C_{LA}^{q})\delta_{A2}^{q}\}\nonumber \\
&&~~~~~~~~\,~+\{(C_{LP}^{q}h^{q}+h^{q}C_{RP}^{q})\delta_{P2}^{q}(C_{RP}^{q}h^{q}+h^{q}C_{LP}^{q})\bar{\delta}_{P2}^{q}\nonumber \\
&&~~~~~~~~~~~~\,~+(C_{LA}^{q}{\tilde h}^{q}+{\tilde h}^{q}C_{RA}^{q})\delta_{A2}^{q}(C_{RA}^{q}{\tilde h}^{q}+{\tilde h}^{q}C_{LA}^{q})\bar{\delta}_{A2}^{q}\}]\nonumber \\
&&+2[tr\{\eta^{l}_{P1}(h^{l})^{2}+\eta^{l}_{A1}(\tilde{h}^{l})^{2}\}tr\{\eta_{P2}^{l}(C_{P}^{l})^{2}+\eta_{A2}^{l}(C_{A}^{l})^{2}\}\nonumber \\
&&~~~\,~+tr\{\eta_{P1}^{l}(C_{P}^{l})^{2}+\eta_{A1}^{l}(C_{A}^{l})^{2}\}tr\{\eta^{l}_{P2}(h^{l})^{2}+\eta^{l}_{A2}(\tilde{h}^{l})^{2}\}]\nonumber \\
&&+\frac{1}{2}[tr\{\eta^{q}_{LP1}(h^{q})^{2}+\eta^{q}_{LA1}(\tilde{h}^{q})^{2}+\eta^{q}_{RP1}(h^{q})^{2}+\eta^{q}_{RA1}(\tilde{h}^{q})^{2}\}\nonumber \\
&&~~~\,~\cdot tr\{\eta_{LP2}^{q}(C_{LP}^{q})^{2}+\eta_{LA2}^{q}(C_{LA}^{q})^{2}+\eta_{RP2}^{q}(C_{RP}^{q})^{2}+\eta_{RA2}^{q}(C_{RA}^{q})^{2}\}\nonumber \\
&&~~~~~+tr\{\eta^{q}_{LP2}(h^{q})^{2}+\eta^{q}_{LA2}(\tilde{h}^{q})^{2}+\eta^{q}_{RP2}(h^{q})^{2}+\eta^{q}_{RA2}(\tilde{h}^{q})^{2}\}\nonumber \\
&&~~~~~\,~\cdot tr\{\eta_{LP1}^{q}(C_{LP}^{q})^{2}+\eta_{LA1}^{q}(C_{LA}^{q})^{2}+\eta_{RP1}^{q}(C_{RP}^{q})^{2}+\eta_{RA1}^{q}(C_{RA}^{q})^{2}\}]\nonumber \\
&&+tr\{\eta^{l}_{P3}(h^{l})^{2}+\eta^{l}_{A3}(\tilde{h}^{l})^{2}\}\nonumber \\
&&~\,\cdot tr\{\eta_{LP3}^{q}(C_{LP}^{q})^{2}+\eta_{LA3}^{q}(C_{LA}^{q})^{2}+\eta_{RP3}^{q}(C_{RP}^{q})^{2}+\eta_{RA3}^{q}(C_{RA}^{q})^{2}\}\nonumber \\
&&+tr\{\eta_{P3}^{l}(C_{P}^{l})^{2}+\eta_{A3}^{l}(C_{A}^{l})^{2}\}\nonumber \\
&&~\,\cdot tr\{\eta^{q}_{LP3}(h^{q})^{2}+\eta^{q}_{LA3}(\tilde{h}^{q})^{2}+\eta^{q}_{RP3}(h^{q})^{2}+\eta^{q}_{RA3}(\tilde{h}^{q})^{2}\}\nonumber \\
&&+\frac{1}{2}tr\{\eta_{P4}^{l}(h^{l})^{2}+\eta_{A4}^{l}({\tilde h}^{l})^{2}\}\nonumber
\\
&&+tr\{\eta_{LP4}^{q}(h^{q})^{2}+\eta_{LA4}^{q}({\tilde h}^{q})^{2}+\eta_{RP4}^{q}(h^{q})^{2}+\eta_{RA4}^{q}({\tilde h}^{q})^{2}\},\label{mu 1}\\
-\mu_{3}^{2}&=&-\frac{3}{16}\alpha tr\{(C^{l}_{P})^{2}\bar{\delta}_{P1}^{l}f^{\dag}f\delta_{P1}^{l}+(C^{l}_{P})^{2}\delta_{P1}^{l}f^{\dag}f\bar{\delta}_{P1}^{l}\nonumber \\
&&~~~~~~~~~~~\,~+(C^{l}_{A})^{2}\bar{\delta}_{A1}^{l}ff^{\dag}\delta_{A1}^{l}+(C^{l}_{A})^{2}\delta_{A1}^{l}ff^{\dag}\bar{\delta}_{A1}^{l}\nonumber \\
&&~~~~~~~~~~~\,~+(f^{\dag}C^{l}_{A}+C^{l}_{P}f^{\dag})\bar{\delta}_{A1}^{l}(C^{l}_{A}f+fC_{P}^{l})\delta_{P1}^{l}\nonumber \\
&&~~~~~~~~~~~\,~+(f^{\dag}C^{l}_{A}+C^{l}_{P}f^{\dag})\delta_{A1}^{l}(C^{l}_{A}f+fC_{P}^{l})\bar{\delta}_{P1}^{l}\}\nonumber \\
&&-\frac{9}{4}\kappa tr\{(C^{l}_{P})^{2}\bar{\delta}_{P2}^{l}f^{\dag}f\delta_{P2}^{l}+(C^{l}_{P})^{2}\delta_{P2}^{l}f^{\dag}f\bar{\delta}_{P2}^{l}\nonumber \\
&&~~~~~~~~~\,~+(C^{l}_{A})^{2}\bar{\delta}_{A2}^{l}ff^{\dag}\delta_{A2}^{l}+(C^{l}_{A})^{2}\delta_{A2}^{l}ff^{\dag}\bar{\delta}_{A2}^{l}\nonumber \\
&&~~~~~~~~~\,~+(f^{\dag}C^{l}_{A}+C^{l}_{P}f^{\dag})\bar{\delta}_{A2}^{l}(C^{l}_{A}f+fC_{P}^{l})\delta_{P2}^{l}\nonumber \\
&&~~~~~~~~~\,~+(f^{\dag}C^{l}_{A}+C^{l}_{P}f^{\dag})\delta_{A2}^{l}(C^{l}_{A}f+fC_{P}^{l})\bar{\delta}_{P2}^{l}\}\nonumber \\
&&+tr(\eta_{P1}^{l}f^{\dag}f+\eta_{A1}^{l}ff^{\dag})tr\{\eta_{P2}^{l}(C_{P}^{l})^{2}+\eta_{A2}^{l}(C_{A}^{l})^{2}\}\nonumber \\
&&+tr\{\eta_{P1}^{l}(C_{P}^{l})^{2}+\eta_{A1}^{l}(C_{A}^{l})^{2}\}tr(\eta_{P2}^{l}f^{\dag}f+\eta_{A2}^{l}ff^{\dag})\nonumber \\
&&+\frac{1}{2}tr(\eta_{P3}^{l}f^{\dag}f+\eta_{A3}^{l}ff^{\dag})\nonumber \\
&&~\,\cdot tr\{\eta_{LP3}^{q}(C_{LP}^{q})^{2}+\eta_{LA3}^{q}(C_{LA}^{q})^{2}+\eta_{RP3}^{q}(C_{RP}^{q})^{2}+\eta_{RA3}^{q}(C_{RA}^{q})^{2}\}\nonumber \\
&&+\frac{1}{4}tr(\eta_{P4}^{l}f^{\dag}f+\eta_{A4}^{l}ff^{\dag}),\\
\lambda_{1}&=&-\frac{3}{8}\alpha tr\{(h^{l})^{2}\bar{\delta}^{l}_{P1}(h^{l})^{2}\delta^{l}_{P1}+({\tilde h}^{l})^{2}\bar{\delta}^{l}_{A1}({\tilde h}^{l})^{2}\delta^{l}_{A1}\nonumber \\
&&~~~~~~~~~\,~+(h^{q})^{2}\bar{\delta}^{q}_{P1}(h^{q})^{2}\delta^{q}_{P1}+({\tilde h}^{q})^{2}\bar{\delta}^{q}_{A1}({\tilde h}^{q})^{2}\delta^{q}_{A1}\}\nonumber \\
&&-\frac{9}{2}\kappa tr\{(h^{l})^{2}\bar{\delta}^{l}_{P2}(h^{l})^{2}\delta^{l}_{P2}+({\tilde h}^{l})^{2}\bar{\delta}^{l}_{A2}({\tilde h}^{l})^{2}\delta^{l}_{A2}\nonumber \\
&&~~~~~~~~~\,~+(h^{q})^{2}\bar{\delta}^{q}_{P2}(h^{q})^{2}\delta^{q}_{P2}+({\tilde h}^{q})^{2}\bar{\delta}^{q}_{A2}({\tilde h}^{q})^{2}\delta^{q}_{A2}\}\nonumber \\
&&+tr\{\eta^{l}_{P1}(h^{l})^{2}+\eta^{l}_{A1}({\tilde h}^{l})^{2}\}tr\{\eta^{l}_{P2}(h^{l})^{2}+\eta^{l}_{A2}({\tilde h}^{l})^{2}\}\nonumber \\
&&+\frac{1}{4}tr\{\eta^{q}_{LP1}(h^{q})^{2}+\eta^{q}_{LA1}({\tilde h}^{q})^{2}+\eta^{q}_{RP1}(h^{q})^{2}+\eta^{q}_{RA1}({\tilde h}^{q})^{2}\}\nonumber \\
&&~\,\cdot tr\{\eta^{q}_{LP2}(h^{q})^{2}+\eta^{q}_{LA2}({\tilde h}^{q})^{2}+\eta^{q}_{RP2}(h^{q})^{2}+\eta^{q}_{RA2}({\tilde h}^{q})^{2}\}\nonumber \\
&&+\frac{1}{2}tr\{\eta^{l}_{P3}(h^{l})^{2}+\eta^{l}_{A3}({\tilde h}^{l})^{2}\}\nonumber \\
&&~\,\cdot tr\{\eta^{q}_{LP3}(h^{q})^{2}+\eta^{q}_{LA3}({\tilde h}^{q})^{2}+\eta^{q}_{RP3}(h^{q})^{2}+\eta^{q}_{RA3}({\tilde h}^{q})^{2}\},\label{lambda1}\\
\lambda_{3}&=&\frac{3}{16}\alpha tr\{(h^{l})^{2}\bar{\delta}^{l}_{P1}(h^{l})^{2}\delta^{l}_{P1}+({\tilde h}^{l})^{2}\bar{\delta}^{l}_{A1}({\tilde h}^{l})^{2}\delta^{l}_{A1}\nonumber \\
&&~~~~~~~~~~+(h^{q})^{2}\bar{\delta}^{q}_{P1}(h^{q})^{2}\delta^{q}_{P1}+({\tilde h}^{q})^{2}\bar{\delta}^{q}_{A1}({\tilde h}^{q})^{2}\delta^{q}_{A1}\}\nonumber \\
&&+\frac{9}{4}\kappa tr\{(h^{l})^{2}\bar{\delta}^{l}_{P2}(h^{l})^{2}\delta^{l}_{P2}+({\tilde h}^{l})^{2}\bar{\delta}^{l}_{A2}({\tilde h}^{l})^{2}\delta^{l}_{A2}\nonumber \\
&&~~~~~~~~~\,~+(h^{q})^{2}\bar{\delta}^{q}_{P2}(h^{q})^{2}\delta^{q}_{P2}+({\tilde h}^{q})^{2}\bar{\delta}^{q}_{A2}({\tilde h}^{q})^{2}\delta^{q}_{A2}\},\\
\rho_{1}&=&-\frac{3}{16}\alpha tr(f^{\dag}f\bar{\delta}_{P1}^{l}f^{\dag}f\delta_{P1}^{l}+ff^{\dag}\bar{\delta}_{A1}^{l}ff^{\dag}\delta_{A1}^{l})\nonumber \\
&&-\frac{9}{4}\kappa tr(f^{\dag}f\bar{\delta}_{P2}^{l}f^{\dag}f\delta_{P2}^{l}+ff^{\dag}\bar{\delta}_{A2}^{l}ff^{\dag}\delta_{A2}^{l})\nonumber \\
&&+\frac{1}{4}tr(\eta_{P1}^{l}f^{\dag}f+\eta_{A1}^{l}ff^{\dag})tr(\eta_{P2}^{l}f^{\dag}f+\eta_{A2}^{l}ff^{\dag}),\label{rho 1}\\
\rho_{2}&=&\frac{3}{32}\alpha tr(f^{\dag}f\bar{\delta}_{P1}^{l}f^{\dag}f\delta_{P1}^{l}+ff^{\dag}\bar{\delta}_{A1}^{l}ff^{\dag}\delta_{A1}^{l})\nonumber \\
&&+\frac{9}{8}\kappa tr(f^{\dag}f\bar{\delta}_{P2}^{l}f^{\dag}f\delta_{P2}^{l}+ff^{\dag}\bar{\delta}_{A2}^{l}ff^{\dag}\delta_{A2}^{l}),\\
\rho_{3}&=&\frac{1}{2}tr(\eta_{P1}^{l}f^{\dag}f+\eta_{A1}^{l}ff^{\dag})tr(\eta_{P2}^{l}f^{\dag}f+\eta_{A2}^{l}ff^{\dag}),\label{rho 3}\\
\alpha_{1}&=&-\frac{3}{16}\alpha tr\{fh^{l}\bar{\delta}_{P1}^{l}h^{l}f^{\dag}\delta_{A1}^{l}+fh^{l}\delta_{P1}^{l}h^{l}f^{\dag}\bar{\delta}_{A1}^{l}\nonumber \\
&&~~~~~~~~~~~\,~+(h^{l})^{2}\bar{\delta}_{P1}^{l}f^{\dag}f\delta_{P1}^{l}+(h^{l})^{2}\delta_{P1}^{l}f^{\dag}f\bar{\delta}_{P1}^{l}\}\nonumber \\
&&-\frac{9}{4}\kappa tr\{fh^{l}\bar{\delta}_{P2}^{l}h^{l}f^{\dag}\delta_{A2}^{l}+fh^{l}\delta_{P2}^{l}h^{l}f^{\dag}\bar{\delta}_{A2}^{l}\nonumber \\
&&~~~~~~~~~\,~+(h^{l})^{2}\bar{\delta}_{P2}^{l}f^{\dag}f\delta_{P2}^{l}+(h^{l})^{2}\delta_{P2}^{l}f^{\dag}f\bar{\delta}_{P2}^{l}\}\nonumber \\
&&+\frac{1}{2}tr(\eta_{P1}^{l}f^{\dag}f+\eta_{A1}^{l}ff^{\dag})tr\{\eta_{P2}^{l}(h^{l})^{2}+\eta_{A2}^{l}({\tilde h}^{l})^{2}\}\nonumber \\
&&+\frac{1}{2}tr\{\eta_{P1}^{l}(h^{l})^{2}+\eta_{A1}^{l}({\tilde h}^{l})^{2}\}tr(\eta_{P2}^{l}f^{\dag}f+\eta_{A2}^{l}ff^{\dag})\nonumber \\
&&+\frac{1}{4}tr(\eta_{P3}^{l}f^{\dag}f+\eta_{A3}^{l}ff^{\dag})\nonumber \\
&&~\,\cdot tr\{\eta_{LP3}^{q}(h^{q})^{2}+\eta_{LA3}^{q}({\tilde h}^{q})^{2}+\eta_{RP3}^{q}(h^{q})^{2}+\eta_{RA3}^{q}({\tilde h}^{q})^{2}\},\label{alpha1}\\
\alpha_{3}&=&\frac{3}{16}\alpha tr\{(h^{l})^{2}\bar{\delta}_{P1}^{l}f^{\dag}f\delta_{P1}^{l}+(h^{l})^{2}\delta_{P1}^{l}f^{\dag}f\bar{\delta}_{P1}^{l}\nonumber \\
&&~~~~~~~~~~+fh^{l}\bar{\delta}_{P1}^{l}h^{l}f^{\dag}\delta_{A1}^{l}+fh^{l}\delta_{P1}^{l}h^{l}f^{\dag}\bar{\delta}_{A1}^{l}\nonumber \\
&&~~~~~~~~~~-(\tilde{h}^{l})^{2}\bar{\delta}_{A1}^{l}ff^{\dag}\delta_{A1}^{l}-(\tilde{h}^{l})^{2}\delta_{A1}^{l}ff^{\dag}\bar{\delta}_{A1}^{l}\nonumber \\
&&~~~~~~~~~~-f^{\dag}\tilde{h}^{l}\bar{\delta}_{A1}^{l}\tilde{h}^{l}f\delta_{P1}^{l}-f^{\dag}\tilde{h}^{l}\delta_{A1}^{l}\tilde{h}^{l}f\bar{\delta}_{P1}^{l}\}\nonumber \\
&&+\frac{9}{4}\kappa tr\{(h^{l})^{2}\bar{\delta}_{P2}^{l}f^{\dag}f\delta_{P2}^{l}+(h^{l})^{2}\delta_{P2}^{l}f^{\dag}f\bar{\delta}_{P2}^{l}\nonumber \\
&&~~~~~~~~~\,~+fh^{l}\bar{\delta}_{P2}^{l}h^{l}f^{\dag}\delta_{A2}^{l}+fh^{l}\delta_{P2}^{l}h^{l}f^{\dag}\bar{\delta}_{A2}^{l}\nonumber \\
&&~~~~~~~~~\,~-(\tilde{h}^{l})^{2}\bar{\delta}_{A2}^{l}ff^{\dag}\delta_{A2}^{l}-(\tilde{h}^{l})^{2}\delta_{A2}^{l}ff^{\dag}\bar{\delta}_{A2}^{l}\nonumber \\
&&~~~~~~~~~\,~-f^{\dag}\tilde{h}^{l}\bar{\delta}_{A2}^{l}\tilde{h}^{l}f\delta_{P2}^{l}-f^{\dag}\tilde{h}^{l}\delta_{A2}^{l}\tilde{h}^{l}f\bar{\delta}_{P2}^{l}\},\label{alpha3}\\
\beta_{1}&=&-\frac{3}{16}\alpha tr(fh^{l}\delta_{P1}^{l}f^{\dag}{\tilde h}^{l}\bar{\delta}_{A1}^{l}+fh^{l}\bar{\delta}_{P1}^{l}f^{\dag}{\tilde h}^{l}\delta_{A1}^{l}\nonumber \\
&&~~~~~~~~~~~~+{\tilde h}^{l}f\bar{\delta}_{P1}^{l}h^{l}f^{\dag}\delta_{A1}^{l}+{\tilde h}^{l}f\delta_{P1}^{l}h^{l}f^{\dag}\bar{\delta}_{A1}^{l})\nonumber \\
&&-\frac{9}{4}\kappa tr(fh^{l}\delta_{P2}^{l}f^{\dag}{\tilde h}^{l}\bar{\delta}_{A2}^{l}+fh^{l}\bar{\delta}_{P2}^{l}f^{\dag}{\tilde h}^{l}\delta_{A2}^{l}\nonumber \\
&&~~~~~~~~~\,~+{\tilde h}^{l}f\bar{\delta}_{P2}^{l}h^{l}f^{\dag}\delta_{A2}^{l}+{\tilde h}^{l}f\delta_{P2}^{l}h^{l}f^{\dag}\bar{\delta}_{A2}^{l}).\label{beta 1}
\end{eqnarray}
\section{Restrictions Derived from Eqs.~($2\cdot$23) and (2$\cdot$25)}
Due to the expressions of the Higgs self-coupling constants in Appendix B, Eqs.~(\ref{1-1}) $\sim$ (\ref{1-4}) lead to four restrictions on several Yukawa coupling constants and several parameters introduced in ${\cal L}_{B}$. The restrictions derived from Eqs.~(\ref{1-2}) and (\ref{1-4}) have been analyzed in \S 3 and 4. Here we give the forms obtained by substituting the expressions of the Higgs self-coupling constants in Appendix B into Eqs.~(\ref{1-1}) and (\ref{1-3}), respectively:
\begin{eqnarray}
&&\alpha (\frac{9}{4}[(C^{l})^{2}\delta^{l}_{1}\bar{\delta}^{l}_{1}tr\{(h^{l})^{2}+(\tilde{h}^{l})^{2}\}+(C^{q})^{2}\delta^{q}_{1}\bar{\delta}^{q}_{1}tr\{(h^{q})^{2}+(\tilde{h}^{q})^{2}\}]\nonumber \\
&&~~~~~~~+\frac{3}{8}\frac{v_{L}^{2}+v_{R}^{2}}{\epsilon_{-}^{2}}\delta^{l}_{1}\bar{\delta}^{l}_{1}[\epsilon_{1}^{2}tr\{(h^{l})^{2}f^{\dag}f\}-\epsilon_{2}^{2}tr\{(\tilde{h}^{l})^{2}ff^{\dag}\}]\nonumber \\
&&~~~+\frac{3}{8}\epsilon_{+}^{2}[\delta^{l}_{1}\bar{\delta}^{l}_{1}tr\{(h^{l})^{4}+(\tilde{h}^{l})^{4}\}+\delta^{q}_{1}\bar{\delta}^{q}_{1}tr\{(h^{l})^{4}+(\tilde{h}^{l})^{4}\}])\nonumber \\
&&+12\kappa (\frac{9}{4}[(C^{l})^{2}\delta^{l}_{2}\bar{\delta}^{l}_{2}tr\{(h^{l})^{2}+(\tilde{h}^{q})^{2}\}+(C^{q})^{2}\delta^{q}_{2}\bar{\delta}^{q}_{2}tr\{(h^{q})^{2}+(\tilde{h}^{l})^{2}\}]\nonumber \\
&&~~~~~~~~+\frac{3}{8}\frac{v_{L}^{2}+v_{R}^{2}}{\epsilon_{-}^{2}}\delta^{l}_{2}\bar{\delta}^{l}_{2}[\epsilon_{1}^{2}tr\{(h^{l})^{2}f^{\dag}f)\}-\epsilon_{2}^{2}tr\{(\tilde{h}^{l})^{2}ff^{\dag})\}]\nonumber \\
&&~~~~~~~~+\frac{3}{8}\epsilon_{+}^{2}[\delta^{l}_{2}\bar{\delta}^{l}_{2}tr\{(h^{l})^{4}+(\tilde{h}^{l})^{4}\}+\delta^{q}_{2}\bar{\delta}^{q}_{2}tr\{(h^{l})^{4}+(\tilde{h}^{l})^{4}\}])\nonumber \\
&=&24\eta_{1}^{l}\eta_{2}^{l}[(C^{l})^{2}tr\{(h^{l})^{2}+(\tilde{h}^{l})^{2}\}+(C^{q})^{2}tr\{(h^{q})^{2}+(\tilde{h}^{q})^{2}\}]\nonumber \\
&&+12\eta_{3}^{l}\eta_{3}^{q}[(C^{q})^{2}tr\{(h^{l})^{2}+(\tilde{h}^{l})^{2}\}+(C^{l})^{2}tr\{(h^{q})^{2}+(\tilde{h}^{q})^{2}\}]\nonumber \\
&&+\frac{1}{2}\eta_{4}^{l}tr\{(h^{l})^{2}+(\tilde{h}^{l})^{2}\}+\frac{1}{2}\eta_{4}^{q}tr\{(h^{q})^{2}+(\tilde{h}^{q})^{2}\}\nonumber \\
&&+\frac{1}{2}(v_{L}^{2}+v_{R}^{2})tr(f^{\dag}f)[2\eta_{1}^{l}\eta_{2}^{l}tr\{(h^{l})^{2}+(\tilde{h}^{l})^{2}\}+\eta_{3}^{l}\eta_{3}^{q}tr\{(h^{l})^{2}+(\tilde{h}^{l})^{2}\}]\nonumber \\
&&+\epsilon_{+}^{2}(\eta_{1}^{l}\eta_{2}^{l}[tr\{(h^{l})^{2}+(\tilde{h}^{l})^{2}\}]^{2}+\eta_{1}^{q}\eta_{2}^{q}[tr\{(h^{q})^{2}+(\tilde{h}^{q})^{2}\}]^{2}\nonumber \\
&&~~~~~\,~+\eta_{3}^{l}\eta_{3}^{q}tr\{(h^{l})^{2}+(\tilde{h}^{l})^{2}\}\cdot tr\{(h^{q})^{2}+(\tilde{h}^{q})^{2}\})\label{1deriv-3}
\end{eqnarray}
and
\begin{eqnarray}
&&(\alpha \delta^{l}_{1}\bar{\delta}^{l}_{1}-12\kappa \delta^{l}_{2}\bar{\delta}^{l}_{2})[\frac{9}{4}(C^{l})^{2}tr(f^{\dag}f)+\frac{3}{8}\epsilon_{1}^{2}tr\{(h^{l})^{2}ff^{\dag}\}\nonumber \\
&&~~~~~~~~~~~~~~~~~~~~~~~~+\frac{3}{8}tr\{(\tilde{h}^{l})^{2}ff^{\dag}\}+\frac{3}{8}(v_{L}^{2}+v_{R}^{2})tr\{(f^{\dag}f)^{2}\}]\nonumber \\
&=&\{24(C^{l})^{2}\eta_{1}^{l}\eta_{2}^{l}+12(C^{q})^{2}\eta_{3}^{l}\eta_{3}^{q}+\frac{1}{2}\eta_{4}^{l}\}tr(f^{\dag}f)\nonumber \\
&&+\frac{1}{2}\epsilon_{+}^{2}[2\eta_{1}^{l}\eta_{2}^{l}tr(f^{\dag}f)tr\{(h^{l})^{2}+(\tilde{h}^{l})^{2}\}+\eta_{3}^{l}\eta_{3}^{q}tr(f^{\dag}f)tr\{(h^{q})^{2}+(\tilde{h}^{q})^{2}\}]\nonumber \\
&&+(v_{L}^{2}+v_{R}^{2})\eta_{1}^{l}\eta_{2}^{l}\{tr(f^{\dag}f)\}^{2}.\label{1deriv-4}
\end{eqnarray}
To make the last form of Eqs.~(\ref{1deriv-3}) and (\ref{1deriv-4}) simple, we have used Eq.~(\ref{tuning-1}) and the following conditions:
\begin{eqnarray}
&&C_{P}^{l ij}=C_{A}^{l ij}\equiv C^{l}\delta^{ij},\\
&&C_{LP}^{q ij}=C_{LA}^{q ij}=C_{RP}^{q ij}=C_{RA}^{q ij}\equiv C^{q}\delta^{ij}.
\end{eqnarray}


\begin{thebibliography}{99}
%
%
\bibitem{1}
I.~S.~Sogami, Prog.~Theor.~Phys.\ {\bf 94} (1995), 117.
\bibitem{2}
K.~Morita, Prog.~Theor.~Phys.\ {\bf 94} (1995), 125.\\
See also:\\
K.~Morita, Y.~Okumura and M.~T.~Yamawaki, Prog.~Theor.~Phys.\ {\bf 94} (1995), 445.
\\
Y.~Okumura, S.~Suzuki and K.~Morita, Prog.~Theor.~Phys.\ {\bf 94} (1995), 1121.
\\
K.~Morita and Y.~Okumura, Prog.~Theor.~Phys.\ {\bf 95} (1996), 227.
\\
Y.~Okumura, Prog.~Theor.~Phys.\ {\bf 95} (1996), 969.
\\
K.~Morita, Prog.~Theor.~Phys.\ {\bf 96} (1996), 787.
\bibitem{3}
A.~Connes and J.~Lott, Nucl.~Phys.\ {\bf B} (Proc.~Suppl.) {\bf 18} (1990), 29.
\\
A.~Connes, in {\it The Interface of Mathematics and Particle Physics}, ed. D.~Quillen, \\G.~B.~Segal and Tsou S.~T. (Clarendon Press, Oxford, 1990).
\\
See also:
\\
A.~H.~Chamseddine and A.~Connes, Phys.~Rev.~Lett.\ {\bf 77} (1996), 4868.
\bibitem{4}
R.~Coquereaux, G.~Esposito-Farese and G.~Vaillant, Nucl.~Phys.\ {\bf B353} (1991), 689.
\\
A.~H.~Chamseddine, G.~Felder and J.~Fr\"ohlich, Phys.~Lett.\ {\bf B296} (1992), 109; Nucl.~Phys.\ {\bf B395} (1993), 672.
\\
A.~Sitarz, Phys.~Lett.\ {\bf B308} (1993), 311.
\\
H.~G.~Ding, H.~Y.~Guo, J.~M.~Li and K.~Wu, Z.~Phys.\ {\bf C64} (1994), 521.
\\
K.~Morita, Prog.~Theor.~Phys.\ {\bf 90} (1993), 219.
\\
K.~Morita and Y.~Okumura, Prog.~Theor.~Phys.\ {\bf 91} (1994), 959; Phys.~Rev.\ {\bf D50} (1994), 1016.
\\
S.~Naka and E.~Umezawa, Prog.~Theor.~Phys.\ {\bf 92} (1994), 189.
\\
J.~Iizuka, H, Kase, K.~Morita, Y.~Okumura and M.~T.~Yamawaki, Prog.~Theor.~Phys.\ {\bf 92} (1994), 397.
\\
Y.~Okumura, Prog.~Theor.~Phys.\ {\bf 92} (1994), 625.
\\
G.~Konisi and T.~Saito, Prog.~Theor.~Phys.\ {\bf 93} (1995), 1093; {\bf 95} (1996), 657.
\\
For applications of the non-commutative geometric method to some L-R symmetric models, see also:
\\
B.~Chen and K.~Wu, AS-ITP-93-64.
\\
B.~E.~Hanlon and G.~C.~Joshi, Lett.~Math.~Phys.\ {\bf 27} (1993), 105.
\\
D.~S.~Hwang and T.~Lee, Int.~J.~Mod.~Phys.\ {\bf A9} (1994), 5531.
\\
B.~Iochum and T.~Sch\"ucker, Lett.~Math.~Phys.\ {\bf 32} (1994), 153.
\\
B.~Chen, H.~G.~Ding, S.~M.~Fei and K.~Wu, Phys.~Lett.\ {\bf B350} (1995), 58.
\\
H.~B.~Benaoum, Mod.~Phys.~Lett.\ {\bf A10} (1995), 479.
\\
K.~Morita and Y.~Okumura, Prog.~Theor.~Phys.\ {\bf 93} (1995), 969.
\\
Y.~Okumura, Prog.~Theor.~Phys.\ {\bf 94} (1995), 589; 607.
\bibitem{5}
I.~S.~Sogami, Prog.~Theor.~Phys.\ {\bf 95} (1996), 637.
\bibitem{6}
N.~G.~Deshpande, J.~F.~Gunion, B.~Kayser and F.~Olness, Phys.~Rev.\ {\bf D44} (1991), 837.\bibitem{7}
R.~N.~Mohapatra and P.~Pal, {\it Massive Neutrinos in Physics and Astrophysics} (World Scientific, Lecture Notes in Physics {\bf 41}, 1991).
\bibitem{double}
J.~F.~Gunion, J.~Grifols, A.~Mendez, B.~Kayser and F.~Olness, Phys.~Rev.\ {\bf D40} (1989), 1546.
\bibitem{8}
J.~Sato, Phys.~Rev.\ {\bf D53} (1996), 3884; Prog.~Theor.~Phys.\ {\bf 96} (1996), 597.
\\
A.~Y.~Smirnov, Nucl.~Phys.\ {\bf B466} (1996), 25.
\bibitem{FCNC}
G.~Ecker, W.~Grimus and H.~Neufeld, Phys.~Lett.\ {\bf B127} (1983), 365.
\\
R.~N.~Mohapatra, G.~Senjanovi\'c and M.~D.~Tran, Phys.~Rev.\ {\bf D28} (1983), 546.
\bibitem{9}
E.~\'Alvarez, J.~M.~Gracia-Bond\'ia and C.~P.~Mart\'in, Phys.~Lett.\ {\bf B306} (1993), 55; {\bf B329} (1994), 259; {\bf B364} (1995), 33.
\\
T.~Shinohara, K.~Nishida, H.~Tanaka and I.~S.~Sogami, Prog.~Theor.~Phys.\ {\bf 96} (1996), 1179.
\\
Y.~Okumura, hep-th/9608208.
\\
For quantum corrections in some models related to the non-commutative geometric method, see also:
\\
A.~H.~Chamseddine and J.~Fr\"ohlich, Phys.~Lett.\ {\bf B314} (1993), 308.
\end{thebibliography}
\end{document}